\documentclass[useAMS, usegraphicx, usenatbib]{mn2e}
\usepackage{aas_macros}
\usepackage{amsmath}
\usepackage{graphicx}
\usepackage{natbib}
\newcommand{\kunit}{\,h\,\mathrm{Mpc}^{-1}}
\newcommand{\runit}{\,h^{-1}\,\mathrm{Mpc}}

\newcommand{\munit}{\,h^{-1}\,\mathrm{M}_{\sun}}
\newcommand{\munitnoh}{\,\mathrm{M}_{\sun}}
\newcommand{\vunit}{\,\mathrm{km}\,\mathrm{s}^{-1}}
\setcitestyle{authoryear,round,comma,aysep={},yysep={,},notesep={}}
\title[Constraining galaxy formation through clustering]{The galaxy correlation function as a constraint on galaxy formation physics}
\author[M. P. van Daalen et al.]{Marcel P. van Daalen$^{1,2,3}$\thanks{E-mail: marcel@berkeley.edu}, Bruno M. B. Henriques$^{1}$, Raul E. Angulo$^{4}$ and
\newauthor Simon D. M. White$^{1}$\\\\
$^1$Max Planck Institute for Astrophysics, Karl-Schwarzschild Stra\ss{}e 1, 85741 Garching, Germany\\
$^2$Leiden Observatory, Leiden University, P.O. Box 9513, 2300 RA Leiden, The Netherlands\\
$^3$Department of Astronomy, Theoretical Astrophysics Center, and Lawrence Berkeley National Laboratory,\\
\phantom{$^3$}University of California, Berkeley, CA 94720, USA\\
$^4$Centro de Estudios de F\'{i}sica del Cosmos de Arag\'{o}n, Plaza San Juan 1, Planta-2, 44001, Teruel, Spain}
\begin{document}
\pagerange{\pageref{firstpage}--\pageref{lastpage}} \pubyear{2015}
\maketitle
\label{firstpage}
\begin{abstract}
We introduce methods which allow observed galaxy clustering to be used together with observed luminosity or stellar mass functions to constrain the physics of galaxy formation. We show how the projected two-point correlation function of galaxies in a large semi-analytic simulation can be estimated to better than $\sim 10\%$ using only a very small subsample of the subhalo merger trees. This allows measured correlations to be used as constraints in a Monte Carlo Markov Chain exploration of the astrophysical and cosmological parameter space. An important part of our scheme is an analytic profile which captures the simulated satellite distribution extremely well out to several halo virial radii. This is essential to reproduce the correlation properties of the full simulation at intermediate separations. As a first application, we use low-redshift clustering and abundance measurements to constrain a recent version of the Munich semi-analytic model. The preferred values of most parameters are consistent with those found previously, with significantly improved constraints and somewhat shifted "best" values for parameters that primarily affect spatial distributions. Our methods allow multi-epoch data on galaxy clustering and abundance to be used as joint constraints on galaxy formation. This may lead to significant constraints on cosmological parameters even after marginalising over galaxy formation physics.
\end{abstract}
\begin{keywords}
galaxies: formation -- cosmology: theory -- cosmology: large-scale structure of Universe
\end{keywords}

\section{Introduction}
Galaxy formation is a complex process involving many astrophysical ingredients spanning a very wide range of scales. Because of this, any model of galaxy formation -- be it hydrodynamical, analytical or semi-analytical in nature -- has to rely on some set of observations in order to constrain the representation of physical processes that cannot be derived from first principles, or be simulated directly.

Hydrodynamical simulations can simulate baryonic processes directly on large scales while relying on sub-grid recipes to model relevant processes below the resolution limit. As such simulations are relatively expensive computationally, the values of the parameters in the sub-grid formulations usually have to be chosen by comparing a set of simulations run at lower resolution or in smaller volumes to some observational quantity, though these numerical settings themselves may impact which parameter values are appropriate. Still, as the available computational resources are ever growing, the number of processes which cannot be simulated directly is slowly decreasing \citep[e.g.][]{Hopkins2014}, and valiant efforts are currently being made to improve the accuracy of direct cosmological simulations (e.g.\ Illustris, \citealt{Vogelsberger2014}, and EAGLE, \citealt{Schaye2015,Crain2015}).

Semi-analytic models (hereafter SAMs), on the other hand, necessarily require calibration of additional physical parameters to describe the baryonic processes which are not simulated directly on any scale. However, once the high-resolution collisionless simulations that they are based on have been run and stored, they can be carried out many times with different parameter values at relatively low computational cost. Coupled with a method to efficiently explore parameter space such as Monte Carlo Markov Chains (MCMC, a method introduced in SAMs by \citealp{Kampakoglou2008} and \citealp{Henriques2009}), this allows one to find the highest-likelihood set of parameters for any given model, based on a set of observational constraints.

Typically, SAMs use observational data sets of one-point functions, such as stellar mass or luminosity functions, as constraints for their model parameters (e.g.\ \citealt{Kauffmann1993,Baugh1996,SomervillePrimack1998,Kauffmann1999,Cole2000,Croton2006,Bower2006,Monaco2007,Somerville2008,Henriques2009,Guo2011,Henriques2013}, see \citealt{Baugh2006} for a review on the general methodology). The resulting models of galaxy formation can then be tested against other observables (i.e.\ observables that are independent of those used as constraints) and be used to make predictions for these. A delicate balance must be maintained here: if the model has too many free parameters, prior regions that are too wide, or if there are too few (independent) observational constraints, degeneracies may occur (i.e.\ wide regions of high likelihood in parameter space, possibly with multiple peaks), while too little freedom or failing to include some relevant physical process may leave the model unable to match several observables at once.

SAMs constrained to match observed luminosity or stellar mass functions have typically had trouble matching the small-scale clustering of galaxies (e.g. \citealt{Kauffmann1999,Springel2005a,Li2007,Guo2011,Kang2012}; but see e.g. \citealt{Kang2014,Campbell2015}). In order to determine the cause of this discrepancy, and to test whether the models retain enough freedom to match the observed clustering at all, it would be instructive to use clustering measurements as additional constraints while exploring parameter space. As galaxy clustering is determined by how galaxies with different properties populate haloes of different mass, it directly constrains galaxy formation, in a way that is complementary to, for example, the luminosity function.

However, this presents a problem: while one-point functions such as the stellar mass function can be quickly estimated with known uncertainty by running the model on a small sample of representative haloes, allowing large regions of parameter space to be rejected without having to run the model on the full dark matter simulation, the same cannot be done simply for two-point functions such as the correlation function. In principle, any observable that relies on spatial correlations between galaxies can only be calculated by running the model on the full simulation, which is computationally infeasible when thousands of parameter sets need to be explored. While running the SAM on a small sub-volume may allow one to measure small-scale correlations to some degree, cosmic variance will be an issue. Additionally, if one aims to compare to observations, where clustering is viewed in projection (unless line-of-sight velocities are used), one still has to account for large-scale correlations, even at small separations.

Here, we present an efficient method, based on the halo model, to estimate the projected correlation function, $w(r_\mathrm{p})$, to some known uncertainty from a small sample of haloes, and we apply it to constrain the version of the Munich semi-analytic model presented in \citet[][, hereafter G13]{Guo2013}. By measuring the properties of galaxies within individual haloes and making informed assumptions about the distribution of these haloes, we are able to circumvent the aforementioned problems, greatly reducing the CPU time needed to predict their two-point clustering.

This paper is organised as follows. In Section~\ref{sec:methods}, we present our method for estimating $w(r_\mathrm{p})$ and briefly describe the semi-analytic model we apply it to. Next, in Section~\ref{sec:results}, we show the results of using clustering as an additional constraint on parameter space, in addition to the oft-used $z=0$ stellar mass function. Finally, in Section~\ref{sec:summary} we present a summary of our work and discuss future improvements and applications.

\section{Method}
\label{sec:methods}

\subsection{Estimating the correlation function}
\label{subsec:estimator}
Our approach is slightly different to that of most previous works constructing a correlation function estimator based on the halo model, where the aim is typically to reproduce observations given some parametrised halo occupation distribution (HOD). Here, our goal is instead to reproduce the results of the semi-analytic model run on the full dark matter simulation to within some given accuracy, given the galaxy properties for a small sample of haloes. As we will show, we are able to reproduce the projected correlation function of the full simulated galaxy sample to within $\sim 10\%$, using the properties of semi-analytical galaxies occupying less than $0.04\%$ of the full halo sample ($0.14\%$ of the subhalo sample).

\subsubsection{The backbone of the model}
\label{subsubsec:backbone}
Our starting point is the linear halo model, introduced independently by \citet{Seljak2000}, \citet{MaFry2000} and \citet{PeacockSmith2000}. In what follows, we will adhere to the terminology of \citet{CooraySheth2002}. In the analytical halo model the power spectrum, $P(k)$, is written as the sum of two terms:
\begin{equation}
P(k)=P^\mathrm{1h}(k)+P^\mathrm{2h}(k).
\label{eq:power}
\end{equation}
Here $P^\mathrm{1h}(k)$ is the 1-halo term, describing the two-point clustering contribution of points within the same halo, and $P^\mathrm{2h}(k)$ is the 2-halo term, describing the contribution of points within separate haloes. For the clustering of matter, these are given by:
\begin{eqnarray}
\nonumber
P_\mathrm{dm}^\mathrm{1h}(k) \!\!\!&=&\!\!\! \int n(M) \left(\frac{M}{\bar{\rho}}\right)^2 |u(k|M)|^2 \mathrm{d}M\\
P_\mathrm{dm}^\mathrm{2h}(k) \!\!\!&=&\!\! \int\!\!\int n(M_1) \left(\frac{M_1}{\bar{\rho}}\right) u(k|M_1) \times\\
\nonumber
& &\!\!\! n(M_2) \left(\frac{M_2}{\bar{\rho}}\right) u(k|M_2) P_\mathrm{hh}(k|M_1,M_2) \mathrm{d}M_1 \mathrm{d}M_2.
\end{eqnarray}
Here $M=M_\mathrm{200mean}$ is the halo mass definition\footnote{$M_\mathrm{200mean}$ is the mass within a spherical region with radius $R_\mathrm{200mean}$ and internal density $200\times \bar{\rho}=200\times \Omega_\mathrm{m}\rho_\mathrm{crit}$.} we will be using throughout, $n(M)$ is the halo mass function, $\bar{\rho}$ is the mean matter density of the Universe, $u(k|M)$ is the normalised Fourier transform of the density profile of a halo of mass $M$, and $P_\mathrm{hh}(k|M_1,M_2)$ is the halo-halo power contributed by two haloes of masses $M_1$ and $M_2$ on a Fourier scale $k$. We can rewrite the latter term assuming a linear scale-independent bias relation, $P_\mathrm{hh}(k|M_1,M_2)=b(M_1)b(M_2)P_\mathrm{lin}(k)$, where $b(M)$ is the halo bias and $P_\mathrm{lin}$ the linear theory matter power spectrum. We then obtain:
\begin{equation}
P_\mathrm{dm}^\mathrm{2h}(k) = P_\mathrm{lin}(k)\left[\int n(M) b(M) \left(\frac{M}{\bar{\rho}}\right) u(k|M) \mathrm{d}M\right]^2\!\!\!\!.
\end{equation}
From these expressions, one can straightforwardly derive a simple model for the galaxy power spectrum. Since we are interested in the clustering of galaxies instead of mass, we replace $M/\bar{\rho}$ by $\left<N_\mathrm{gal}|M\right>/\bar{n}_\mathrm{gal}$ and, since galaxies are discrete objects, $\left(M/\bar{\rho}\right)^2$ by $\left<N_\mathrm{gal}(N_\mathrm{gal}-1)|M\right>/\bar{n}_\mathrm{gal}^2$, leading to:
\begin{eqnarray}
P_\mathrm{gal}^\mathrm{1h}(k) \!\!\!&=&\!\!\! \int n(M) \frac{\left<N_\mathrm{gal}(N_\mathrm{gal}-1)|M\right>}{\bar{n}_\mathrm{gal}^2} u_\mathrm{gal}(k|M)^p \mathrm{d}M\\
\nonumber
P_\mathrm{gal}^\mathrm{2h}(k) \!\!\!&=&\!\!\! P_\mathrm{lin}(k)\left[\int n(M) b(M) \frac{\left<N_\mathrm{gal}|M\right>}{\bar{n}_\mathrm{gal}} u_\mathrm{gal}(k|M) \mathrm{d}M\right]^2\!\!\!\!.
\end{eqnarray}
Here the mean number density of galaxies is given by $n_\mathrm{gal}=\int n(M) \left<N_\mathrm{gal}|M\right>\mathrm{d}M$. Note that we have followed \citet{CooraySheth2002} in replacing the normalised Fourier transform of the halo density profile, $u(k|M)$, by one describing the distribution of (satellite) galaxies, $u_\mathrm{gal}(k|M)$, and subsequently in changing the power-law index on this term in the 1-halo term by $p$. This is often done in the literature in order to be able to differentiate between contributions from central-satellite and satellite-satellite terms, with $p=1$ for the former and $p=2$ for the latter, based on the value of $\left<N_\mathrm{gal}(N_\mathrm{gal}-1)\right>$. $\left<N_\mathrm{gal}|M\right>$ -- the most common form of the HOD -- is often separated into contributions from centrals and satellites as well, with the former ($N_\mathrm{cen}$) being either $0$ or $1$, and the latter ($N_\mathrm{sat}$) being very well approximated by a (linear) power law \citep[e.g.][]{GuzikSeljak2002,Kravtsov2004,Zehavi2005,Tinker2005,Zheng2005}. From this, approximate expressions for $\left<N_\mathrm{gal}(N_\mathrm{gal}-1)\right>$ in terms of $N_\mathrm{cen}$ and $N_\mathrm{sat}$ can be derived as well.

However, as our aim is to reproduce the results of the semi-analytic model, for which information on the HOD and the galaxy type is much more readily available than for observations, we can explicitly separate the contributions from central and satellite galaxies to the galaxy power spectrum without approximation. Keeping in mind that a halo will contain at most one central, meaning that $\left<N_\mathrm{cen}(N_\mathrm{cen}-1)|M\right>=0$, that $\left<N_\mathrm{cen}N_\mathrm{sat}|M\right>=\left<N_\mathrm{sat}N_\mathrm{cen}|M\right>$, and using that central galaxies reside in the centre of the halo and should therefore not be weighted by the profile, we derive:
\begin{eqnarray}
\nonumber
P_\mathrm{gal}^\mathrm{1h}(k) \!\!\!\!\!\!&=&\!\!\!\!\!\! 2\!\!\int\!\! n(M) \frac{\left<N_\mathrm{cen}N_\mathrm{sat}|M\right>}{\bar{n}_\mathrm{gal}^2} \left[u_\mathrm{gal}(k|M) - W_{R}(k)\right] \mathrm{d}M\, +\\
\nonumber
& & \!\!\!\!\!\!\!\int\!\! n(M) \frac{\left<N_\mathrm{sat}(N_\mathrm{sat}\!\!-\!1)|M\right>}{\bar{n}_\mathrm{gal}^2} \!\left[u_\mathrm{gal}(k|M)^2 \!\!-\! W_{R}(k)^2\right]\! \mathrm{d}M\\
P_\mathrm{gal}^\mathrm{2h}(k) \!\!\!\!\!\!&=&\!\!\!\!\!\! P_\mathrm{lin}(k)\left[\int\!\! n(M) b(M) \frac{\left<N_\mathrm{cen}|M\right>}{\bar{n}_\mathrm{gal}} \mathrm{d}M\, +\right.\\
\nonumber
& & \left.\!\!\!\!\!\!\!\int\!\! n(M) b(M) \frac{\left<N_\mathrm{sat}|M\right>}{\bar{n}_\mathrm{gal}} u_\mathrm{gal}(k|M) \mathrm{d}M\right]^2\!\!\!\!.
\label{eq:haloterms}
\end{eqnarray}
Note that we have followed \citet{ValageasNishimichi2011} in adding a counterterm to the halo profiles in the 1-halo term, which ensures the 1-halo term goes to zero for $k \rightarrow 0$. Here $W_{R}(k)$ is the Fourier transform of a spherical top-hat of radius $R(M)=[3M/(4\pi\bar{\rho})]^{1/3}$, given by:
\begin{equation}
W_{R}(k) = 3\left(\frac{\sin(kR)}{(kR)^3}-\frac{\cos(kR)}{(kR)^2}\right).
\end{equation}

In our model, we take $P_\mathrm{lin}(k)$ to be the measured power spectrum of the initial conditions of the dark matter simulation. We calculate the halo mass function, $n(M)$, directly from the dark matter simulation too and spline-fit the results. Furthermore, we use the fit for the $M_\mathrm{200mean}$ halo bias function provided by \citet{Tinker2010} for $b(M)$, and compute each of the four HOD terms directly from the SAM run on our halo subsample, spline-fitting these results as well.

\subsubsection{The galaxy distribution}
The normalised Fourier transform of the galaxy distribution, $u_\mathrm{gal}(k|M)$, is often derived from the dark matter mass profile of the halo. This in turn is usually assumed to be equal to the \citet[][, NFW]{Navarro1997} profile, cut off at the virial radius $r_\mathrm{vir}=R_\mathrm{200mean}$, with some concentration-mass relation $c(M)$:
\begin{equation}
\rho_\mathrm{NFW}(r)=\frac{\rho_0}{(r/r_\mathrm{s})(1+r/r_\mathrm{s})^2},
\end{equation}
where $r_\mathrm{s}=r_\mathrm{vir}/c$ is the scale radius. The main advantage of using the one-parameter NFW profile is that this leads to an analytic expression for $u(k|M)$. However, many authors have shown that the \citet{Einasto1965} profile provides a more accurate fit to the mean profile of haloes of a given mass, and to the distribution of dark matter substructure \citep[e.g.][]{Navarro2004,Merritt2005,Merritt2006,Gao2008,Springel2008,Stadel2009,Navarro2010,Reed2011,DuttonMaccio2014}. The two-parameter Einasto density profile is given by:
\begin{equation}
\rho_\mathrm{Ein}(r)=\rho_0 \exp\left\{-\frac{2}{\alpha}\left[\left(\frac{r}{r_\mathrm{s}}\right)^{\alpha}-1\right]\right\},
\end{equation}
where the shape parameter $\alpha$ allows additional freedom in the slope of the profile. This function does not have an analytic Fourier transform, and an extra numerical integration step is therefore needed when replacing the NFW profile by an Einasto one. The larger degeneracies in fitting a two-parameter model also mean more data points are needed to obtain a reliable fit. Still, when the computational expense is acceptable and enough information on the measured profile is available, the increased accuracy will be worth the cost.

We find that the Einasto profile provides a very good fit to the distribution of satellite galaxies in the inner parts of haloes in our simulation. But even the Einasto profile over-predicts the number of galaxies at large radii, $r \ga 0.7r_\mathrm{vir}$. Furthermore, standard practice is to cut off the profile at the virial radius, while we find that $\sim 10\%$ of the satellite galaxies in our simulation are found at distances $1<r/r_\mathrm{vir}<3$. Note that these galaxies are not necessarily outside the virialised region, as haloes are typically not spherical objects. In addition, simulated haloes are truncated in a non-spherical manner at the boundary imposed by the Friends-of-Friends (FoF) group finder used to define them. Finally, subhaloes may travel outside the virial radius again after infall \citep[``backsplash'', e.g.][]{Balogh2000,Mamon2004,Gill2005}. We therefore seek a profile with similar small-scale behaviour to the Einasto profile, while simultaneously fitting the galaxy population of simulated haloes out to $\sim 3 r_\mathrm{vir}$.

We find that the following functional form is capable of providing an excellent match to the galaxy distribution over the full range of scales we consider, and at any halo mass:
\begin{equation}
n_\mathrm{sat}(r)=n_0 \left(\frac{r}{b}\right)^{a-3}\exp\left\{-\left(\frac{r}{b}\right)^c\right\}.
\label{eq:preprofile}
\end{equation}
This fitting function has three parameters, $a$, $b$ and $c$. Note that the role of $b$ is similar to that of $r_\mathrm{s}$ in the Einasto profile. Both the Einasto and this new profile are near universal if defined in terms of $x \equiv r/r_\mathrm{vir}$. If we rewrite both profiles in terms of $x$ and integrate them to obtain $N(<r)$, the similarities and differences between the profiles are most easily appreciated. For the Einasto profile:
\begin{equation}
N_\mathrm{Ein}(<r) = N_\mathrm{tot} \frac{\gamma\left[\frac{3}{\alpha},\frac{2}{\alpha}\left(\frac{x}{r_\mathrm{s}}\right)^\alpha\right]}{\gamma\left[\frac{3}{\alpha},\frac{2}{\alpha}\left(\frac{x_\mathrm{max}}{r_\mathrm{s}}\right)^\alpha\right]},
\end{equation}
while for the profile given in equation \eqref{eq:preprofile}:
\begin{equation}
N_\mathrm{sat}(<r) = N_\mathrm{tot} \frac{\gamma\left[\frac{a}{c},\left(\frac{x}{b}\right)^c\right]}{\gamma\left[\frac{a}{c},\left(\frac{x_\mathrm{max}}{b}\right)^c\right]}.
\end{equation}
\begin{figure}
\begin{center}
\includegraphics[width=1.0\columnwidth, trim=7mm 8mm 0mm 8mm]{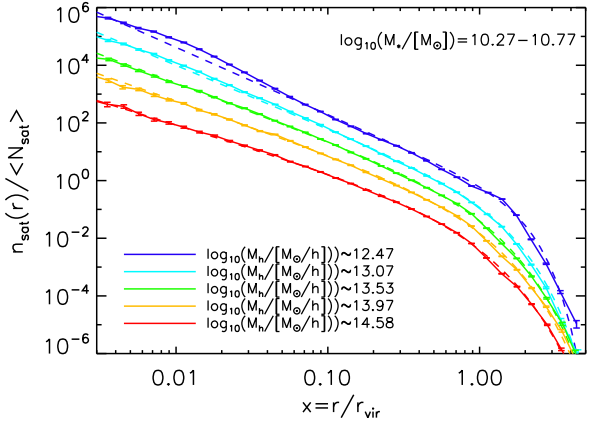}
\caption{Galaxy number density profiles for all \citet{Guo2011} halo members with stellar masses $10.27<\log_{10}(M_*/\mathrm{M}_{\sun})<10.77$, for five different halo mass bins (shown in different colours). The legend shows the mean logarithmic mass in each of the bins. Solid lines indicate the measured profiles, while dashed lines show our best fits (see equation \eqref{eq:profile}). The halo mass bins are dynamically chosen such that each contains roughly the same number of galaxies, and the fits are performed using 30 radial bins spaced equally in log-space between $\log_{10}x=-2.5$ and $\log_{10}x=0.5$. The three-parameter fit we use, given in equation \eqref{eq:profile}, captures the measured satellite profile extremely well, even out to several times the virial radius, where the popular NFW and Einasto profiles often fail because of the edge imposed by halo definition in the simulation. For more information on how the fits are performed, see Appendix~\ref{fittingapp}.}
\label{fig:profiles}
\end{center}
\end{figure}
Here $\gamma(a,b)$ is the lower incomplete gamma function, and we have assumed the profiles cut off at some $x_\mathrm{max}$. The similarities in the two profiles are clear. The main difference is that the two parameters of the gamma function can be independently manipulated for our new profile, which effectively allows for a steeper profile at large $x$ and consequently a better match to the galaxy distribution around the virial radius. In practice, we fit a normalised number density profile $n_\mathrm{sat}(r)/\left<N_\mathrm{sat}\right>$ to the satellite distribution before numerically Fourier transforming this to obtain $u_\mathrm{gal}(k|M)$. For completeness, $n_\mathrm{sat}(r)/\left<N_\mathrm{sat}\right>$ is given by:
\begin{equation}
\label{eq:profile}
\frac{n_\mathrm{sat}(r)}{\left<N_\mathrm{sat}\right>}=\frac{c}{4\pi b^3 r_\mathrm{vir}^3\gamma\left[\frac{a}{c},\left(\frac{x_\mathrm{max}}{b}\right)^c\right]} \left(\frac{x}{b}\right)^{a-3}\exp\left\{-\left(\frac{x}{b}\right)^c\right\}.
\end{equation}
In our model we set $x_\mathrm{max}=5$. Even for small halo samples, the three parameters of the fit are sufficiently independent to ensure degeneracies are not a problem, i.e.\ fits with the new profile provide stable results even for a small number of sample points.

An example is given in Figure~\ref{fig:profiles}, where we show the best-fit model for all halo members with stellar masses $10.27<\log_{10}(M_*/\mathrm{M}_{\sun})<10.77$ in the \citet{Guo2011} semi-analytic model, for five different halo mass bins. The solid lines show the measured number density profiles, while the dashed lines show the best fits for the profile in equation \eqref{eq:profile}. The radii of satellite galaxies in each halo mass bin are not binned when fitting, but instead used as direct inputs for a likelihood function which we maximise to find the best fit. This likelihood function is constructed assuming that the number of satellites found at each radius is a Poisson variable with mean given by the profile (see Appendix~\ref{fittingapp}).

After finding the best-fit profile parameters in each halo mass bin, we fit an Akima spline through each of the three parameters as a function of halo mass to obtain smooth functions that are stable to outliers. Not only does the profile given in equation \eqref{eq:profile} fit the simulated satellite distribution extremely well for a large range in mass and radius, it is also parametrised in a way that yields very stable fits when only a few galaxies are available. This is important as we will only be using a very small set of haloes to inform our model, and the resulting clustering prediction is very sensitive to the satellite profile fits.

\subsubsection{Halo exclusion}
The standard halo model does not account for halo exclusion, meaning that the distance between two haloes is allowed to be arbitrarily close to zero. As a consequence, the 2-halo term is overestimated on small scales. We have implemented halo exclusion following the approach of \citet{Baldauf2013}. They suggest a correction to the 2-halo term, such that $P_\mathrm{2h}'(k)=P_\mathrm{2h}(k)-P_\mathrm{corr}(k)$. If the sum of two halo radii (i.e.\ their minimum separation) is $R$, this correction term is given by $P_\mathrm{corr}(k)=V_\mathrm{excl}\left[P_\mathrm{2h}\ast W_{R}\right](k)+V_\mathrm{excl}W_{R}(k)$, where $[A \ast B](k)$ denotes a convolution integral between functions $A(k)$ and $B(k)$, and where $V_\mathrm{excl}=(4\pi/3)R^3$ is the effective excluded volume. For galaxies, the correction term can be split into separate contributions for central-central, central-satellite and satellite-satellite galaxy pairs, as:
\begin{eqnarray}
\nonumber
P_\mathrm{corr}(k) \!\!\!\!\!\!&=&\!\!\!\!\!\! (1-f_\mathrm{sat})^2 P_\mathrm{corr,cc}(k)+\\
& & \!\!\!\!\!\!2 f_\mathrm{sat}(1-f_\mathrm{sat}) P_\mathrm{corr,cs}(k)+f_\mathrm{sat}^2 P_\mathrm{corr,ss}(k),
\end{eqnarray}
where $f_\mathrm{sat}$ is the satellite fraction. We calculate averaged minimum separation radii $R_\mathrm{cc}$, $R_\mathrm{cs}$ and $R_\mathrm{ss}$ separately for each pairing and each stellar mass bin, by integrating over all halo masses. The final halo exclusion correction terms are then given by:
\begin{eqnarray}
\nonumber
P_\mathrm{corr,cc}(k) \!\!\!\!\!\!&=&\!\!\!\!\!\! V_\mathrm{excl}\left[P_\mathrm{2h,cc}\ast W_{R_\mathrm{cc}}\right](k)+V_\mathrm{excl}W_{R_\mathrm{cc}}(k),\\
\nonumber
P_\mathrm{corr,cs}(k) \!\!\!\!\!\!&=&\!\!\!\!\!\! \left(u_\mathrm{gal}(k|M_\mathrm{h,sat})-W_{R_\mathrm{sat}}(k)\right)\times\\
\label{eq:haloexclusion}
& & \!\!\!\!\!\!\left(V_\mathrm{excl}\left[P_\mathrm{2h,cs}\ast W_{R_\mathrm{cs}}\right](k)+V_\mathrm{excl}W_{R_\mathrm{cs}}(k)\right),\\
\nonumber
P_\mathrm{corr,ss}(k) \!\!\!\!\!\!&=&\!\!\!\!\!\! \left(u_\mathrm{gal}(k|M_\mathrm{h,sat})^2-W_{R_\mathrm{sat}}(k)^2\right)\times\\
\nonumber
& & \!\!\!\!\!\!\left(V_\mathrm{excl}\left[P_\mathrm{2h,ss}\ast W_{R_\mathrm{ss}}\right](k)+V_\mathrm{excl}W_{R_\mathrm{ss}}(k)\right),
\end{eqnarray}
where $M_\mathrm{h,sat}$ is the typical halo mass of the satellite population. Note that we have again taken into account the counterterms proposed by \citet{ValageasNishimichi2011}, with $R_\mathrm{sat}$ corresponding to the size of the Lagrangian region of the typical halo mass of the satellite population.

Implementing halo exclusion significantly improves our model predictions around the 1-halo and 2-halo transition scale ($r\sim 1\,\mathrm{Mpc}$) for our most massive stellar mass bin. For our fiducial model, we observed no noticeable effects on the projected correlation function for stellar masses $M_* \la 10^{11}\munitnoh$.

\begin{figure}
\begin{center}
\includegraphics[width=1.0\columnwidth, trim=9mm 8mm 0mm 7mm]{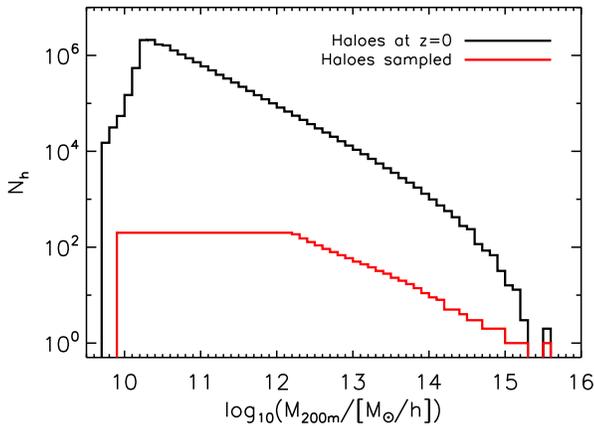}
\caption{The FoF halo mass function, showing the number of haloes available in the Millennium Simulation at $z=0$ (black) and the number randomly selected as a function of $M_\mathrm{200mean}$ in each subsample (red). The subsamples each comprise less than $0.04\%$ of the total halo sample, or $0.14\%$ of the total subhalo sample. The selection function was built iteratively by demanding that $\sim 99\%$ of the random samples it generates lead to projected correlation functions that are within $30\%$ of the full sample prediction. Low-mass haloes were favoured over high-mass haloes in order to suppress the size of the trees used in the SAM. Even so, the fraction of FoF groups needed to match the correlation function within some uncertainty at any stellar mass is higher for more massive haloes.}
\label{fig:selection}
\end{center}
\end{figure}

\subsubsection{Correction for non-sphericity and halo alignment}
As is common, we have assumed a spherical distribution of satellite galaxies around each central. In reality, haloes and consequently their galaxy populations are triaxial. \citet{vanDaalen2012} investigated the effect of assuming a spherical distribution on the two-point correlation function and galaxy power spectrum, and found that the effects can be quite large, with the true power being underestimated by $1\%$ around $k=0.2\kunit$ to $10\%$ around $k=25\kunit$, increasing even more towards smaller scales (see the right panel of their Figure 3). We have repeated their analysis and found that the functional shape of this underestimation of the power appears to be completely independent of the stellar mass of the galaxies. We therefore fit a function $e(k)$ through these results and use this to correct our halo model power spectra for the combined effects of non-sphericity and halo alignment. The final galaxy power spectrum that comes out of our model for a given set of galaxies is therefore:
\begin{equation}
P_\mathrm{gal}=[P_\mathrm{gal}^\mathrm{1h}(k)+P_\mathrm{gal}^\mathrm{2h}(k)]/[1+e(k)],
\label{eq:totgalpow}
\end{equation}
with $P_\mathrm{gal}^\mathrm{1h}(k)$ and $P_\mathrm{gal}^\mathrm{2h}(k)$ given by equation~\eqref{eq:haloterms}.

\begin{figure}
\begin{center}
\includegraphics[width=1.0\columnwidth, trim=7mm 8mm 0mm 8mm]{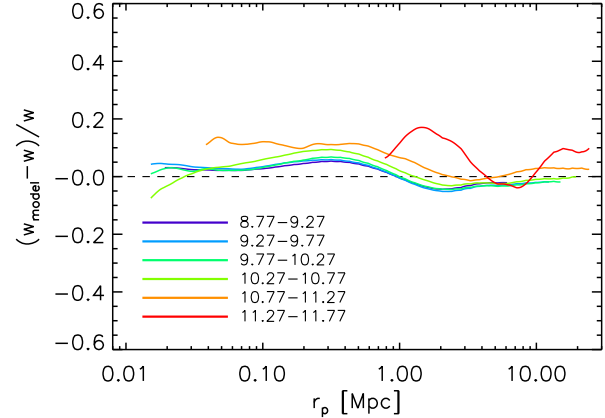}
\caption{The fractional difference between our model predictions of the projected galaxy correlation function and a direct calculation, for galaxies in the \citet{Guo2011} semi-analytic model. Here we use the full simulated galaxy sample as an input to our model. Results are shown for six different stellar mass bins, indicated by lines of different colours, over the range where SDSS/DR7 data is available for each. The overall agreement is within $10\%$. The most massive stellar mass bin is most sensitive to the approximations made in the halo model, and has only a small number of satellites, making it difficult to get an accurate prediction for its (relatively) small-scale behaviour. However, this mass bin also has next to no impact on the constraints on the galaxy formation parameters, as it is only weakly sensitive to them. On the other hand, the four least massive stellar mass bins, which are the most sensitive to the parameters of the SAM, show excellent agreement between the real and predicted clustering functions.}
\label{fig:wguo}
\end{center}
\end{figure}

\begin{figure*}
\begin{center}
\begin{tabular}{ccc}
\includegraphics[width=1.0\columnwidth, trim=7mm 8mm 0mm 8mm]{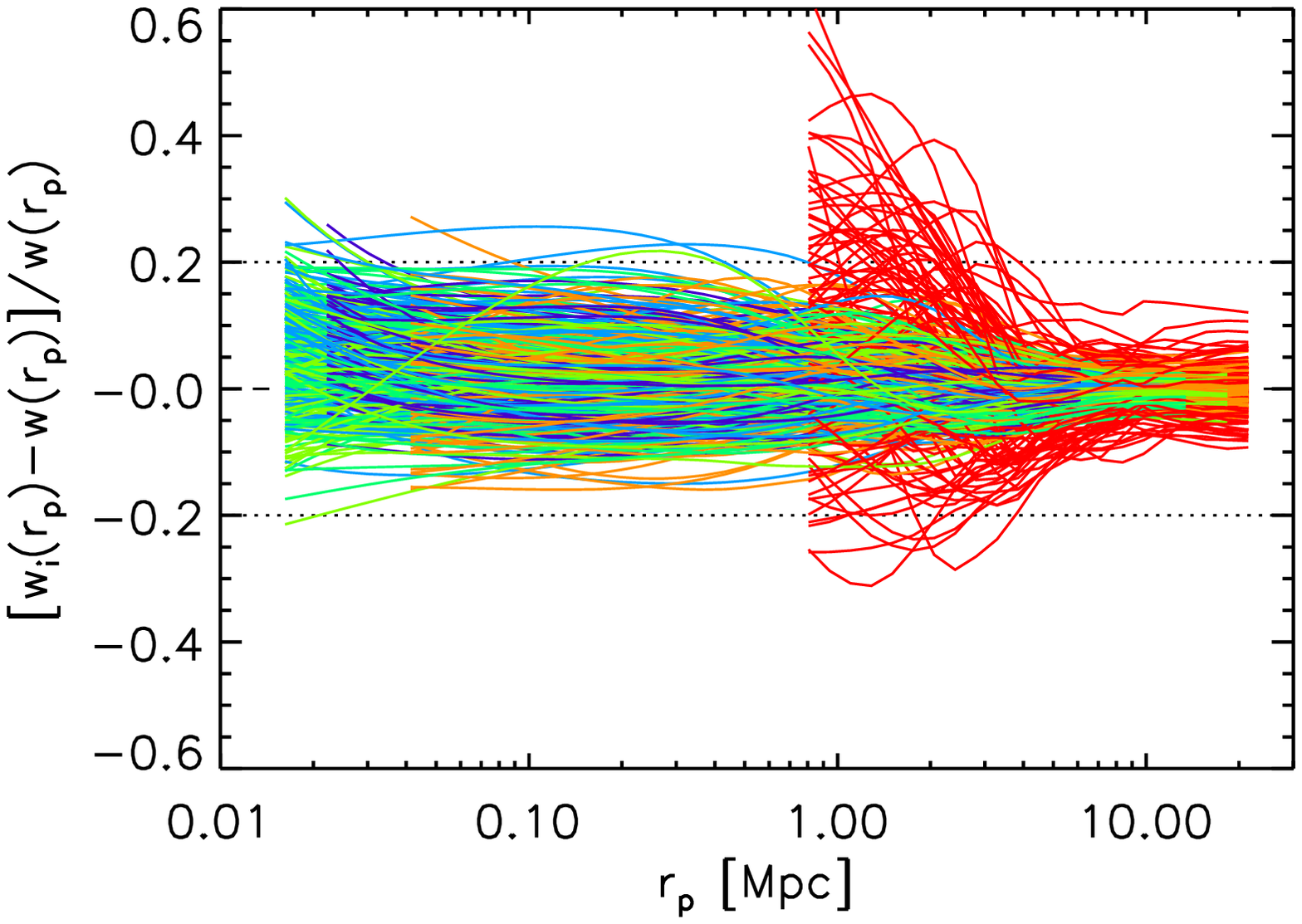} & &
\includegraphics[width=1.0\columnwidth, trim=7mm 8mm 0mm 8mm]{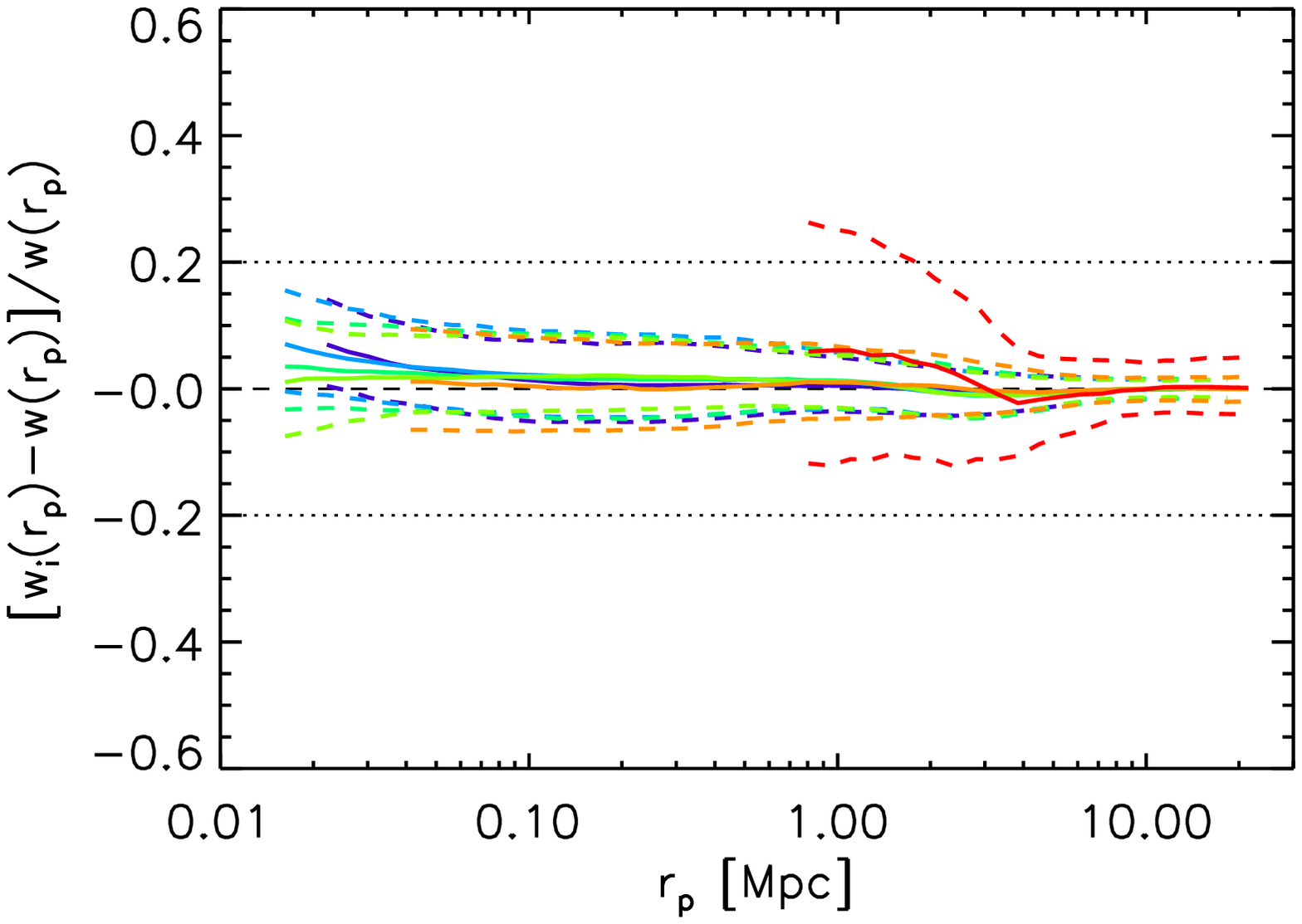}
\end{tabular}
\caption{The fractional difference between the predictions of our model for different halo subsamples and the model prediction for the full sample. Each subsample consists of about $0.14\%$ of the total subhalo sample (see text for details). The colours indicate the same stellar mass bins as in Figure~\ref{fig:wguo}. Dotted lines are shown at $\pm 20\%$ for reference. \textit{Left:} The predictions of each of $100$ separate realisations, showing the scatter around the full sample result. \textit{Right:} Same as the left panel, but now showing only the median, $16^\mathrm{th}$ and $84^\mathrm{th}$ percentiles for $500$ different samples. Even using a small random sample, our model can quickly estimate the projected correlation function to $\sim 10\%$ precision. Note that even more accurate results can be achieved at the cost of a larger halo sample.}
\label{fig:wsamples}
\end{center}
\end{figure*}

\subsubsection{Assembly bias}
\citet{Zentner2014} have shown that not including assembly bias in a clustering model leads to significant systematics. As \citet{vanDaalen2012} showed, assembly bias is also reflected in the galaxies of the \citet{Guo2011} model. However, assembly bias is generally only a significant systematic when the galaxies are split by a property other than mass, e.g.\ colour or age, which we do not do here. Most importantly, since our clustering estimator is based on the HOD and satellite profile of galaxies in the simulation, assembly bias is implicitly included in our results. Indeed, as we will show in \S\ref{subsubsec:performance}, our estimator does not seem to contain any significant scale-independent systematics. We therefore do not include an explicit correction for assembly bias in our model.

\subsubsection{Converting to the projected correlation function}
To obtain the projected correlation function from the galaxy power spectrum, we numerically perform two standard transformations. First, to obtain the 3D correlation function:
\begin{equation}
\xi(r)=\frac{1}{2\pi^2}\int_0^\infty k^2 P(k)\frac{\sin kr}{kr}\,\mathrm{d}k,
\end{equation}
and, finally, to obtain the projected galaxy correlation function:
\begin{equation}
w(r_\mathrm{p})=2\int_0^{\pi_\mathrm{lim}} \!\!\xi\!\left(\!\sqrt{r_\mathrm{p}^2+\pi^2}\right)\mathrm{d}\pi=2\int_{r_\mathrm{p}}^{r_\mathrm{lim}} \!\!\frac{r\xi(r)}{\sqrt{r^2-r_\mathrm{p}^2}}\,\mathrm{d}r.
\end{equation}
Here $r_\mathrm{p}$ and $\pi$ are the projected and line-of-sight separation, respectively, and $r_\mathrm{lim}=\sqrt{r_\mathrm{p}^2+\pi_\mathrm{lim}^2}$. Formally, the integration limit is $r_\mathrm{lim}=\infty$, but in order to directly compare our model $w(r_\mathrm{p})$ to that of observations we set $r_\mathrm{lim}=40\runit$, and convert all units from $\mathrm{Mpc/h}$ to $\mathrm{Mpc}$. As \citet{vandenBosch2013} point out, this ignores the contribution of peculiar velocities beyond the integration limit, which may bias the projected correlation function on the largest scales probed. However, since the largest scales are the least interesting for our current investigation, we do not attempt to correct for this.

\subsubsection{Selection function}
\label{subsubsec:selection}
Sample haloes are randomly selected following the selection function shown in Figure~\ref{fig:selection}, a power law with a cut-off at a maximum of $200$ haloes per $0.1\,\mathrm{dex}$ in halo mass.

The selection function was initially constructed iteratively by demanding that the projected correlation functions resulting from $\ga 99\%$ of its random samples should agree with those from the full sample in each stellar mass bin to within $30\%$. Low-mass haloes were favoured over high-mass haloes by weighting each halo by the average number of subhaloes for its mass, in order to down-weight large merger trees. For the same reason, the constraint on the accuracy of the projected correlation function for galaxies with $M_*>10^{11.27}\munitnoh$ was relaxed, as it would require almost all of the highest-mass haloes to achieve $\sim 10\%$ accuracy consistently for the clustering of these rare (mostly central) galaxies.

After building several selection functions in this way, we found that on average they were well approximated by the combination of a constant value and a power law (rounded to integer values) shown as the red line in Figure~\ref{fig:selection}. The subsamples generated by this selection function each comprise less than $0.04\%$ of the total FoF halo sample, or $0.14\%$ of the total subhalo sample.

\begin{table*}
\begin{center}
\begin{tabular}{lclcccc}
\hline
\hline
Parameter & & Description & & Units\\
\hline
\hline
$\alpha_{\rm{SF}}$ & & Star formation efficiency & & --\\
$\tilde{M}_{\rm{crit}}$ & & Star formation threshold & & $\mathrm{M}_{\sun}\,\mathrm{km}\,\mathrm{s}^{-1}\,\mathrm{Mpc}^{-1}$\\
$\alpha_{\rm{SF,burst}}$ & & Star formation burst mode efficiency & & --\\
$\beta_{\rm{SF,burst}}$ & & Star formation burst mode slope & & --\\
\hline
$k_{\rm{AGN}}$ & & Radio feedback efficiency & & $h^{-1}\,\mathrm{M}_{\sun}\,\mathrm{yr}^{-1}$\\
$f_{\rm{BH}}$ & & Black hole growth efficiency & & --\\
$V_{\rm{BH}}$ & & Quasar growth scale & & $\mathrm{km}\,\mathrm{s}^{-1}$\\
\hline
$\epsilon$ & & SN mass-loading efficiency & & --\\
$V_{\rm{reheat}}$ & & Mass-loading scale & & $\mathrm{km}\,\mathrm{s}^{-1}$\\
$\beta_{1}$ & & Mass-loading slope & & --\\
$\eta$ & & SN ejection efficiency & & --\\
$V_{\rm{eject}}$ & & SN ejection scale & & $\mathrm{km}\,\mathrm{s}^{-1}$\\
$\beta_{2}$ & & SN ejection slope & & --\\
$\gamma$ & & Ejecta reincorporation scale factor & & --\\
\hline
$y$ & & Metal yield fraction & & --\\
\hline
$R_{\rm{merger}}$ & & Major-merger threshold ratio & & --\\
$\alpha_{\rm{friction}}$ & & Dynamical friction scale factor & & --\\
%\hline
%$M_{\rm{ram}}$ & & Ram-pressure stripping threshold halo mass & & $10^{10}\munit$\\
\hline
\hline
\end{tabular}
\end{center}
\caption{Parameters varied in the MCMC. The best-fit values (as well as the G13 values and the prior ranges) are shown in Figure~\ref{fig:params}. For more information we refer to G13.}
\label{tab:params}
\end{table*}

\subsubsection{Performance of the model}
\label{subsubsec:performance}
We compare $w(r_\mathrm{p})$ predicted by applying our estimator to the full halo sample to that measured directly on the simulation for the \citet{Guo2011} model in Figure~\ref{fig:wguo}. Here we show the relative difference between the two for six different bins in stellar mass, indicated as ranges in $\log_{10}(M_*/\mathrm{M}_{\sun})$. We only show the results over the range where we constrain $w(r_\mathrm{p})$ using observations. The model performs well, and deviations from the true correlation function are below $15\%$ except for the most massive galaxy bin. The magnitude of the mismatch tends to increase with stellar mass. The large-scale disagreement is caused by the model slightly under-predicting the power in the transition region between the 1-halo and 2-halo terms, while the small-scale offset is mostly due to the 1-halo term in the power spectrum being slightly overestimated around $k=1\kunit$. However, overall the agreement is good, especially considering our relatively simple treatment of e.g.\ the halo bias (linear and scale-independent), and we leave further improvements -- such as using a halo-halo power spectrum measured from the dark matter only simulation instead of a biased linear power spectrum -- to future work. Note that the clustering predictions in the four lowest mass bins ($M_*<10^{10.27}\munitnoh$) are always accurate to better than $10\%$. This is important, as the clustering of these galaxies on sub-Mpc scales is sensitive to changes in the galaxy formation parameters, and therefore holds the most constraining power.

The true power of the model lies in its ability to reproduce the clustering prediction for the full sample from only a small subsample of FoF groups. In Figure~\ref{fig:wsamples} we compare the predictions for a large set of random subsamples selected according to the selection function shown in Figure~\ref{fig:selection} to the model prediction for the full sample. The dotted lines indicate offsets of $\pm 20\%$ for reference, and the colours indicate the same stellar mass bins as in Figure~\ref{fig:wguo}. For all but the highest mass bin, the model on average reproduces the clustering prediction of the full sample to the $1\%$ level, with a sample-to-sample scatter that is typically $<10\%$. This shows that the model is capable of reproducing the full sample estimate using only a small fraction of all the haloes.

While the model is sensitive to the number of high-mass haloes used, it is not sensitive to the low-mass end. Raising the low-mass ceiling of our selection function from $200$ to $1000$ haloes only slightly lowers the scatter for $r_\mathrm{p}<40\,\mathrm{kpc}$ while having no significant effect on larger scales, while lowering the ceiling to $40$ haloes increases the scatter on all scales by about $50\%$ while increasing the median offset only for $r_\mathrm{p}<40\,\mathrm{kpc}$, to on average $10\%$ on the smallest scales.

We find that the accuracy of the estimator is not sensitive to the galaxy formation parameters used, but is instead mainly determined by the particular haloes in the sample. Indeed, as we will see, our predictor works equally well for all sets of best-fit parameters we explore in \S\ref{sec:results}. This means that we could in principle construct an optimal (i.e.\ maximally representative) sample of haloes, given some selection function, and reasonably expect this sample to give highly accurate predictions throughout the SAM's parameter space. Here, we have instead chosen the simpler approach of generating several random halo samples and using the one that lies closest to the medians shown in Figure~\ref{fig:wsamples}.

\subsection{The SAM and MCMC}
\label{subsec:SAM}
As our estimator is able to quickly and accurately recover the projected correlation function from a very small subsample of haloes, this makes it ideally suited for constraining the parameter space of semi-analytic models using the projected correlation function. In this work we present a first application, where we constrain the model of G13, a recent version of the Munich semi-analytical code, using both the galaxy stellar mass function (SMF) and the projected galaxy correlation function. For this we utilise the same data sets as presented in G13. Since we will only utilise the Millennium Simulation, and not Millennium II, we only use constraints for stellar masses $M_*>10^9\munit$.

\begin{figure*}
\begin{center}
\includegraphics[width=1.0\textwidth, trim=0mm 6mm 0mm 6mm]{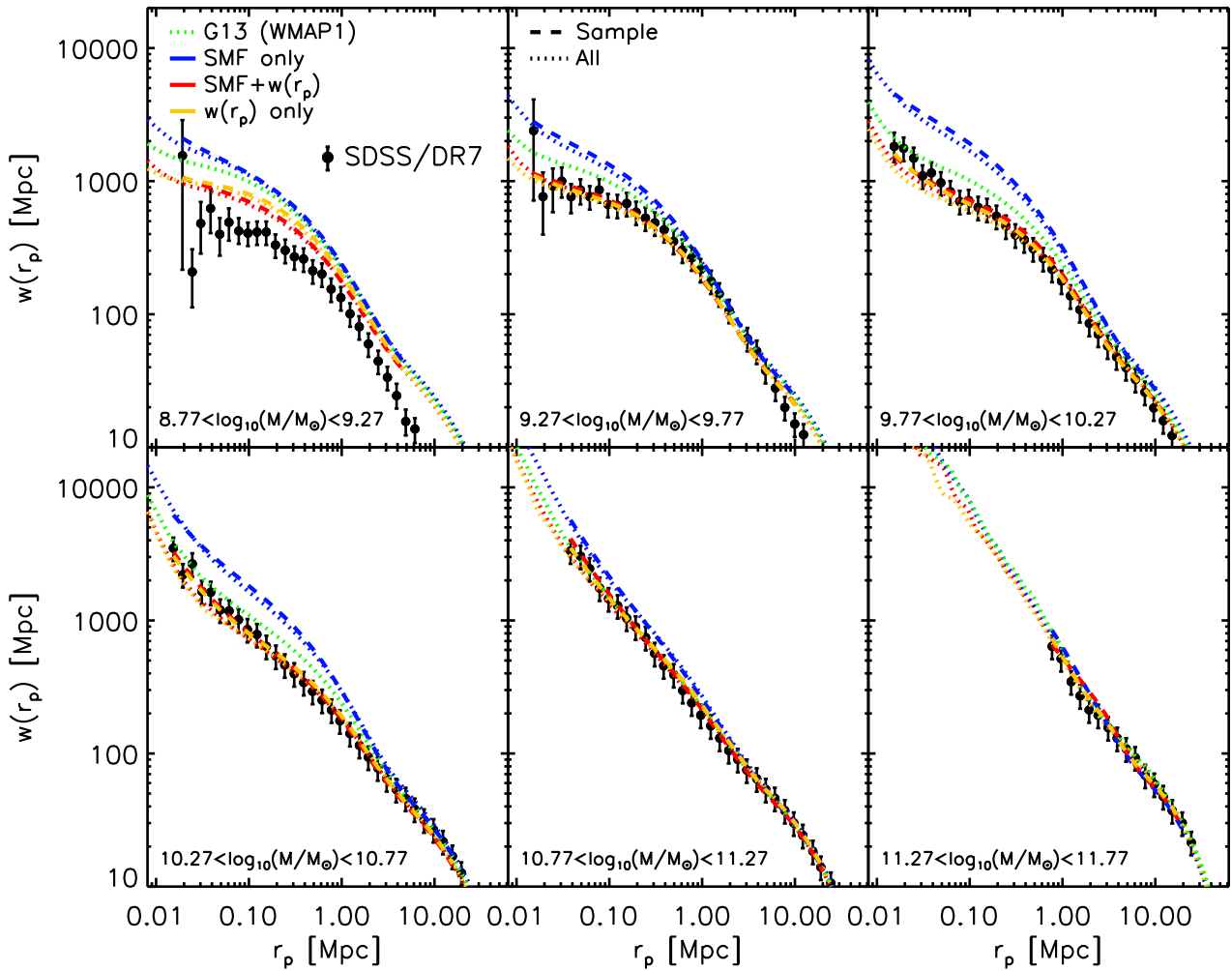}
\caption{The projected galaxy correlation function in six bins of stellar mass. The points with error bars show the SDSS data in each bin, while the lines show the model results. The green dotted line shows the results for the original model from G13, in which the parameter values were set manually. The blue lines show the results of only using the stellar mass function as a constraint, the orange lines show the results of only using the projected correlation functions as constraints, while the red lines show the results when the model is simultaneously constrained by both data sets. Finally, dashed and dotted lines are used to indicate whether these are the results for the sample haloes or for all haloes, respectively. The clustering predicted from the sample agrees with the actual clustering to $\la 10\%$, as expected (see \S\ref{subsubsec:performance}). The $w(r_\mathrm{p})$-only and SMF+$w(r_\mathrm{p})$ correlation functions for the sample are in almost perfect agreement with the SDSS data used as constraints, even though the latter is simultaneously constrained by the stellar mass function. Note that even though the lowest mass bin is not used as a constraint, the match to observations for these two models is markedly improved with respect to the others.}
\label{fig:corr_constrained}
\end{center}
\end{figure*}

The G13 model includes $17$ parameters which together determine the outcome of galaxy formation. These are (see Table~\ref{tab:params}): the star formation efficiency ($\alpha_\mathrm{SF}$); the star formation criterion ($\tilde{M}_\mathrm{crit}$, or equivalently $\Sigma_\mathrm{crit}$); the star formation efficiency in the burst phase following a merger ($\alpha_\mathrm{SF,burst}$); the slope on the merger mass ratio determining the stellar mass formed in the burst ($\beta_\mathrm{SF,burst}$); the AGN radio mode efficiency ($k_\mathrm{AGN}$); the black hole growth efficiency ($f_\mathrm{BH}$); the typical halo virial velocity of the black hole growth process ($V_\mathrm{BH}$); three parameters governing the reheating and injection of cold disk gas into the hot halo phase by supernovae (SNe), namely the gas reheating efficiency ($\epsilon$), the reheating cut-off velocity ($V_\mathrm{reheat}$) and the slope of the reheating dependence on $V_\mathrm{vir}$ ($\beta_1$); three parameters governing the ejection of hot halo gas to an external reservoir, namely the gas ejection efficiency ($\eta$), the ejection cut-off velocity ($V_\mathrm{eject}$) and the slope of the ejection dependence on $V_\mathrm{vir}$ ($\beta_2$); a parameter controlling the gas return time from the external reservoir to the hot halo ($\gamma$); the yield fraction of metals returned to the gas phase by stars ($y$); the mass ratio separating major and minor merger events ($R_\mathrm{merger}$); and finally a parameter controlling the dynamical friction time scale of orphan galaxies, i.e. the time it takes for satellite galaxies of which the dark matter subhalo is disrupted (or at least no longer detected) to merge with the central galaxy ($\alpha_\mathrm{friction}$).

While in the original G13 paper some of these parameters were held fixed, here we allow all $17$ to vary in order to determine which of these are sensitive to the $w(r_\mathrm{p})$ constraints. We start our Monte Carlo Markov Chains (MCMCs) at the position in parameter space preferred by \citet{Guo2011} and the WMAP1 version of G13, which was arrived at by requiring agreement with a variety of low-redshift observational data, primarily stellar mass and luminosity functions, but also gas fractions, gaseous and stellar metallicities, and central black hole masses, all as a function of stellar mass. Here we use MCMC techniques to find new sets of "best-fit" parameters (i.e.\ the parameters that result in the best agreement with the data) constrained by one or both of the low-redshift stellar mass function and the projected correlation function.

Since the error bars on the SDSS clustering data were derived from Poisson statistics alone, and so do not include cosmic variance, we artificially increase them for our fitting. Data points for the observed projected correlation function with uncertainties below $20\%$ had their error bar increased to this minimum value. The SMF was treated as having the same minimum uncertainty in order to avoid skewing our estimates. This is appropriate because we wish our MCMC procedures to exclude regions of parameter space where models substantially mismatch the data, rather than to attempt a statistically rigorous estimate of model parameters. As noted before, we do not use the clustering data below $M_*=10^{9.27}\munitnoh$, nor the stellar mass function data below $M_*<10^9\munitnoh$, when constraining the model, as the haloes hosting these galaxies are not well resolved in the Millennium Simulation which we are using as a basis for the SAM. When fitting to the SMF and clustering data simultaneously, we increase the relative weighting of the fit to the SMF by a factor of five to compensate for the fact that the clustering data is measured in five separate bins. This helps avoid sacrificing the excellent fit to the SMF in favour of matching the correlation function.\footnote{We note that our results are not sensitive to the choice of the weight factor: we found that in our case using a factor of two or ten instead of the fiducial five yielded the exact same set of best-fit parameters.}

Note that while G13 used both WMAP1 and WMAP7 cosmologies, we here use the original WMAP1 cosmology only to avoid additional complications introduced by scaling to a different cosmology. In future work the results will be explored for more up-to-date cosmologies. As G13 showed, the change in cosmology does have some impact on the resulting correlation functions, although they are at least as sensitive to the SAM's physical recipes. Besides updating the cosmology, the only change made from the WMAP1 \citet{Guo2011} model to the newer G13 model is that the type 2 (orphan) satellite galaxy positions are now correctly updated in the code, meaning that their orbits now decay as intended and can therefore be disrupted earlier. This change was the reason for the improved agreement with clustering data in the WMAP1 version of G13 with respect to \citet{Guo2011}.

\begin{figure}
\begin{center}
\includegraphics[width=1.0\columnwidth, trim=8mm 8mm 4mm 6mm]{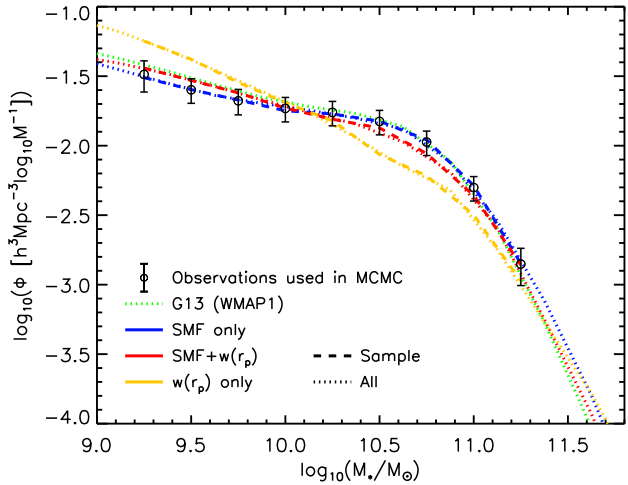}
\caption{The stellar mass functions of the models. The lines are as in Figure~\ref{fig:corr_constrained}, and here the estimates for the sample are in perfect agreement with the full calculation. Even when the SMF is not the only constraint, the model clearly has enough freedom to reproduce it to high precision, as the SMF+$w(r_\mathrm{p})$ results are within 1$\sigma$ of the data. The badly-matched $w(r_\mathrm{p})$-only model shows that reproducing the galaxy clustering does not guarantee reproducing the stellar mass function.}
\label{fig:smf_constrained}
\end{center}
\end{figure}

\section{Results}
\label{sec:results}

\subsection{Comparison with observations}
\label{subsec:comparison}
The results of our MCMC chains for the projected correlation function are shown in Figure~\ref{fig:corr_constrained}, for six bins in stellar mass, as indicated in the panels. In each figure, we indicate the original results found by G13, where the galaxy formation parameters were set by hand without reference to clustering, as a green dotted line. The new results are shown in blue, orange and red; in blue, we show the correlation functions that follow from using the stellar mass function alone as a constraint (``SMF-only''), in orange we show the result of using the clustering data alone as a constraint (``$w(r_\mathrm{p})$-only''), while in red we show the results of constraining with both data sets simultaneously (``SMF+$w(r_\mathrm{p})$''). Note that the correlation functions for $w(r_\mathrm{p})$-only and SMF+$w(r_\mathrm{p})$ tend to coincide.

The dashed lines show the predictions made based on the sample of haloes used in the MCMC, as described in \S\ref{subsec:estimator}. The dotted lines show the true galaxy correlation function, as calculated directly from the full simulated galaxy catalogue for the same model parameters. The true values agree very well with the ones estimated from the sample (at the $\la 10\%$ level), as expected from the results of \S\ref{subsubsec:performance}.

\begin{figure}
\begin{center}
\includegraphics[width=1.0\columnwidth, trim=8mm 8mm 4mm 6mm]{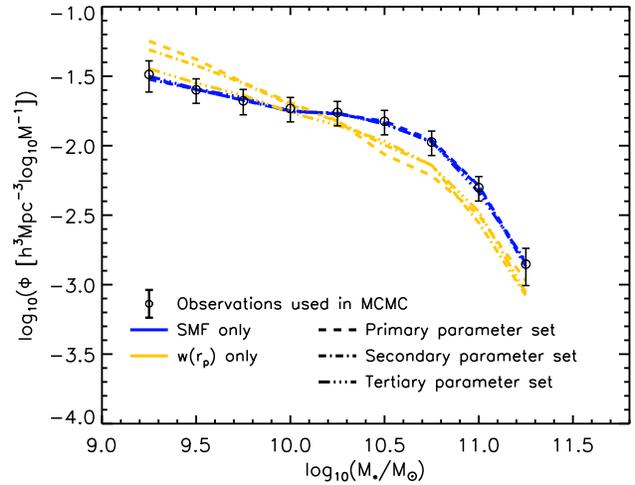}
\caption{Same as Figure~\ref{fig:smf_constrained}, but now showing the results for three sets of parameters instead of only the single best-fit. In each case we show the maximum likelihood model and two other models with only slightly lower likelihood but parameters spread over the allowed regions. Full and SMF+$w(r_\mathrm{p})$ results have been omitted for clarity. While the $w(r_\mathrm{p})$-only model in general prefers a lowered SMF around and above the knee, it does not necessarily increase the number of low-mass galaxies to match the clustering.}
\label{fig:smf_alt}
\end{center}
\end{figure}

The resulting SMF+$w(r_\mathrm{p})$ correlation functions (red lines) provide a better fit to the data, bringing the small-scale clustering down considerably in comparison with the original G13 and SMF-only (blue lines) models. This effect is larger for low stellar masses, where the clustering discrepancy between the old model and the data was larger as well. The much improved match to observations indicates that the model retains enough freedom to match the clustering data. Note that the match to the projected galaxy correlation function for galaxies in the first mass bin is significantly improved as well, even though this data is not used to constrain the model. For the highest-mass galaxies, $11.27 < \log_{10}(M_*/\mathrm{M}_{\sun}) < 11.77$, all models perform equally well, as the clustering of these galaxies is relatively insensitive to the galaxy formation parameters.

In Figure~\ref{fig:smf_constrained}, we show how the models compare to the SMF data used to constrain them. The black points with error bars are derived by combining several observational data sets \citep[see][]{Henriques2015}. The error bars show the uncertainties we used to calculate the likelihoods (see \S\ref{subsec:SAM}). The original G13 model, in which the parameters were set by hand, is again shown as a green dotted line, and matches the data well. When we use only the SMF as a constraint for the galaxy formation model, shown in blue, we obtain a marginally better fit to the data at low mass.

\begin{figure*}
\begin{center}
\includegraphics[width=1.0\textwidth, trim=0mm 6mm 0mm 6mm]{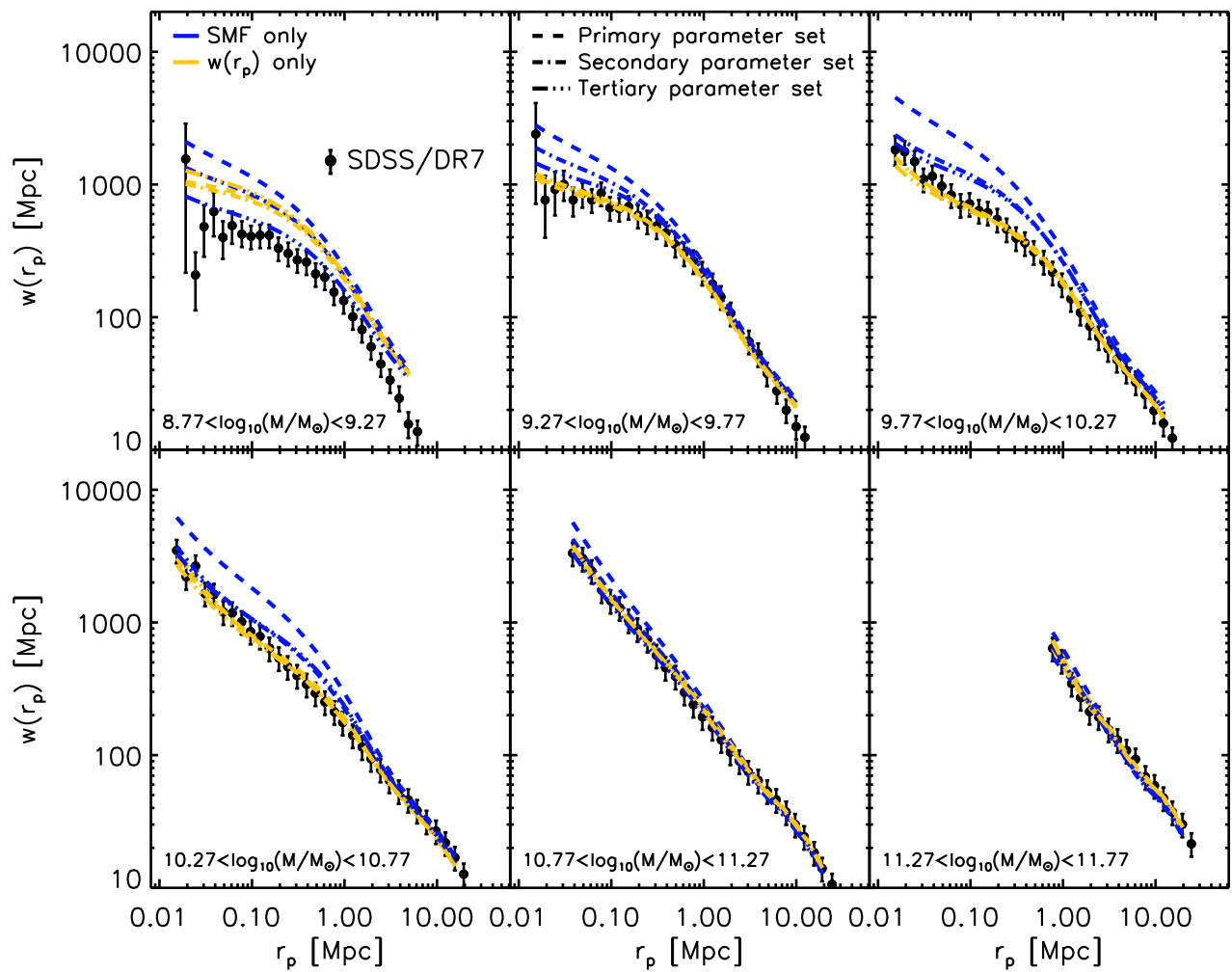}
\caption{Same as Figure~\ref{fig:corr_constrained}, but now showing the results for three sets of parameters instead of only the single best-fit. The parameter sets used here are identical to those of Figure~\ref{fig:smf_alt}. Full and SMF+$w(r_\mathrm{p})$ results have been omitted for clarity. Even though the three high-likelihood parameter sets for $w(r_\mathrm{p})$-only shown here yield nigh-identical projected correlation functions, they can have markedly different stellar mass functions. While the clustering data serve as a powerful constraint, they should not be used instead of the stellar mass function, but as a complement to it. Similarly, the clustering of models that provide perfect fits to the stellar mass function (i.e.\ SMF-only) can be very different.}
\label{fig:corr_alt}
\end{center}
\end{figure*}

When the projected galaxy correlation function is used as an additional constraint, shown in red, the agreement with the stellar mass function is almost as good. The simulated mass function for SMF+$w(r_\mathrm{p})$ shows slightly worse agreement with observations than SMF-only around the knee, where a decrease in clustering tends to necessitate a decrease in galaxy abundance, but the result is still within 1$\sigma$ of the data. The sample results (dashed lines) agree perfectly with the full catalogue ones (dotted lines) for all models, which is expected as the stellar mass function is a one-point function and requires a smaller halo sample to achieve the same accuracy as the correlation function.

It is clear that the SMF+$w(r_\mathrm{p})$ model is in much closer agreement with both the SMF and the clustering data \emph{simultaneously} than both the original G13 and the SMF-only models, as the latter models are in (sometimes strong) disagreement with the clustering data for low-mass galaxies on small scales while the SMF+$w(r_\mathrm{p})$ model is generally in agreement with both the low-mass clustering data and the SMF within 1$\sigma$. This shows the merit of using clustering estimates as constraints while exploring parameter space. In \S\ref{subsec:parameters}, we will show that adding the projected correlation functions as constraints not only markedly improves the constraints on almost all the SAM parameters, but that there is at least one model parameter which is \emph{only} significantly constrained by including clustering data.

Looking at the results for $w(r_\mathrm{p})$-only (orange lines in Figures~\ref{fig:corr_constrained} and \ref{fig:smf_constrained}), we see that its (predicted) projected correlation functions, like those for SMF+$w(r_\mathrm{p})$, are in almost perfect agreement with the data, while its stellar mass function is inconsistent with the observations at the many $\sigma$ level for both small and large stellar masses. It is clearly important to use both the number density and clustering data as constraints.

To show that reproducing the projected correlation functions does not dictate a certain SMF, we show in Figure~\ref{fig:smf_alt} the stellar mass functions resulting from using two other $w(r_\mathrm{p})$-only models with only slightly lower likelihood but parameters spread over the allowed regions. One of these sets results in an SMF that happens to reproduce the observed SMF at the 1$\sigma$ level for $M_*<10^{10.3}\munitnoh$. We find that $w(r_\mathrm{p})$-only models that produce practically identical correlation functions (see Figure~\ref{fig:corr_alt}) can show a wide range of number densities for low-mass galaxies, but will almost always underestimate the stellar mass function at high mass, especially at the knee. We explore the reason for this in the next section.

For good measure we also show results for three high-likelihood sets of parameters for SMF-only in both Figures~\ref{fig:smf_alt} and \ref{fig:corr_alt}; virtually indistinguishable stellar mass functions can produce very different correlation functions, many of them highly incompatible with observations. Our results show that the clustering data, while powerful, should not be used as the only constraint for the model, and -- to perhaps a larger degree -- neither should the $z=0$ stellar mass function be the only observational constraint. We note that the latter point was also made by \citet{Henriques2009}, where large degeneracies were obtained when using only this constraint even when sampling only six parameters.

Because the observations we use here are not sufficient to constrain all model parameters, we do not consider the effect our best-fit parameters have on other observable quantities, but instead leave this to future work where additional constraints (such as high-redshift information) are adopted and/or some model parameters are held fixed.

\subsection{Changes in parameters}
\label{subsec:parameters}
Even though we vary $17$ galaxy formation parameters, by far the largest role in bringing the clustering predictions in agreement with observations is played by only two of these: $\alpha_\mathrm{friction}$, which controls the time it takes for satellite galaxies to merge with the central once their dark matter subhalo has been disrupted, and $V_\mathrm{reheat}$, which indirectly controls the amount of cold ISM gas reheated to the hot halo by supernova feedback as a function of halo mass.

\begin{figure*}
\begin{center}
\includegraphics[width=1.0\textwidth, trim=12mm 10mm 0mm 8mm]{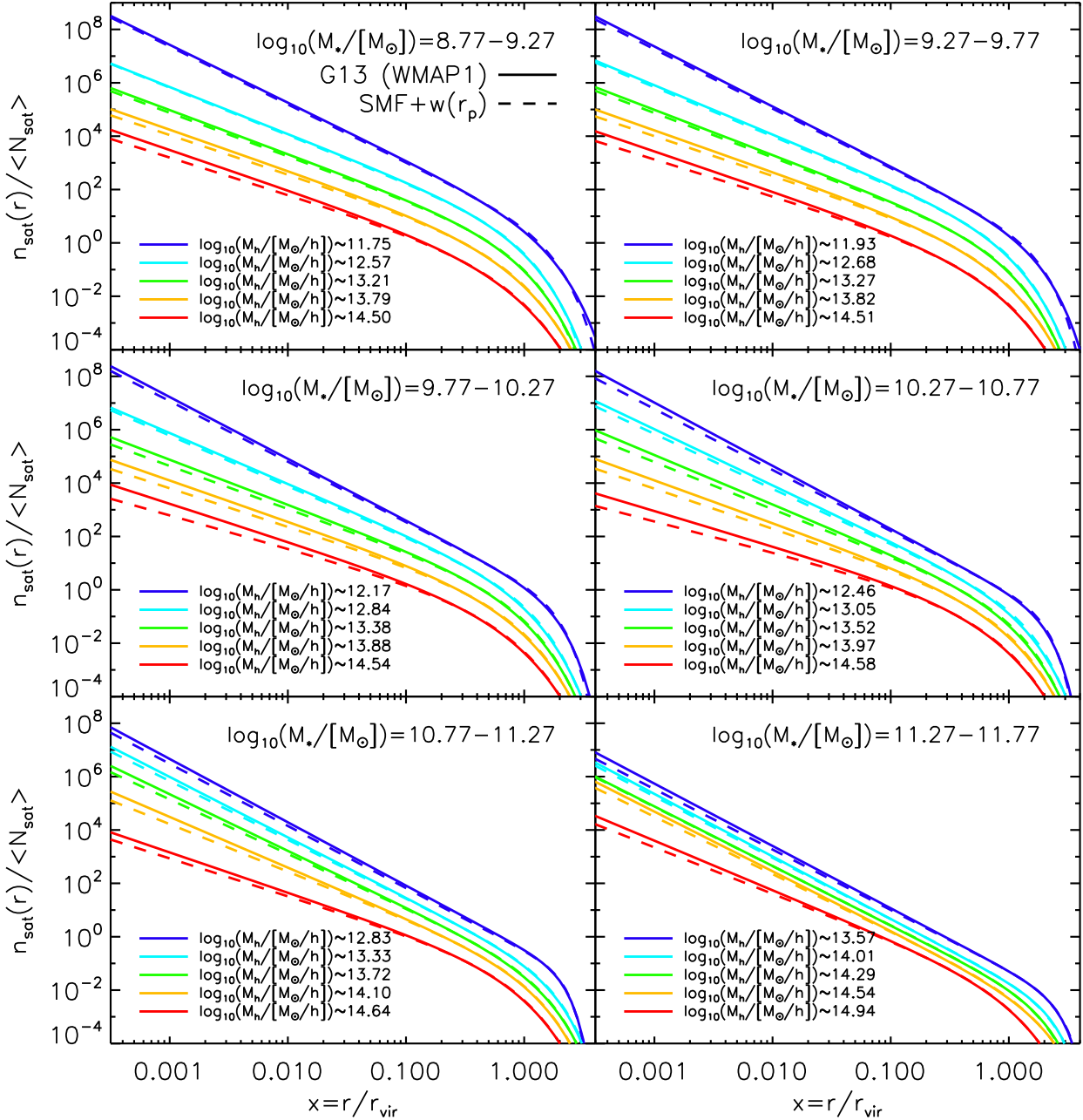}
\caption{Comparison of the galaxy distribution profiles for G13 (solid lines) and the SMF+$w(r_\mathrm{p})$ (dashed lines) best-fit parameters. The different panels show the profiles of galaxies in the six correlation function mass bins, as indicated in the top right of each panel. As in Figure~\ref{fig:profiles}, different colours are used for different halo mass bins, which are set to be the same for both models to allow for an unbiased comparison. Note that we do change the mass bins as a function of stellar mass in order to make sure each bin in halo mass is roughly equally populated. For clarity, we show only the fits to the measured profiles (see equation \ref{eq:profile}) here, but stress that each provides an excellent fit over the full range shown. Note that the dynamic range in scales has been extended relative to Figure~\ref{fig:profiles} to better appreciate the differences between the profiles. Mainly because of the reduced dynamical friction time scale in the latter model, the profiles of galaxies in every mass bin are slightly flatter at any halo mass, reducing the correlation function on small scales. This shows that the correlation functions are sensitive to very small changes in the satellite profile.}
\label{fig:profilecompare}
\end{center}
\end{figure*}

The way these parameters influence the clustering and stellar mass function predictions is as follows. When the clustering data are included as an additional constraint, the dynamical friction time scale of orphan galaxies decreases by about $25\%$ with respect to G13. This small but significant shift causes galaxies at small separation scales to merge with their centrals more quickly, flattening the galaxy distribution profile within the haloes and decreasing the amount of clustering on small scales, especially for low-mass satellites. This change in the galaxy distribution profiles from the G13 to the SMF+$w(r_\mathrm{p})$ model is shown in Figure~\ref{fig:profilecompare}. The halo mass bins are set to be the same for the two models to allow for an unbiased comparison. Note that the mass bins do change as a function of stellar mass in order to make sure each bin in halo mass is roughly equally populated. Although we only show the fits to the measured profiles here, we stress that each provides an excellent fit to the data, over the full range in scales shown. The change in slope of the profiles, caused mainly by the change in $\alpha_\mathrm{friction}$, is relatively small, meaning that the galaxy distributions are still consistent with SDSS data for rich clusters (see Figure 11 of G13).

The $25\%$ decrease in the friction time scale when using clustering as an additional constraint causes the number of type 2 galaxies of any mass at $z=0$ to decrease by a third; however, we note that this roughly one-to-one correspondence between $\alpha_\mathrm{friction}$ and the number of orphan galaxies is coincidental. In general, we find that the dynamical friction timescale can decrease by a factor of a few without lowering the number of satellites further, as the merging time scale for many of these galaxies is still long compared to the Hubble time. While the lower number density of orphans hardly affects the total number density of low-mass galaxies, the small measure of flattening experienced by the profile is enough to significantly decrease the small-scale clustering for these galaxies.

The apparent sensitivity of low-mass clustering to a relatively small change in $\alpha_\mathrm{friction}$, which does not seem to be reflected in another observational quantity -- at least not at current observational precision -- means that using the clustering data as a constraint for semi-analytic models of galaxy formation may be one of the few ways in which the right merging time scale for orphan satellite galaxies can be determined. Whether SAMs use a parametrised dynamical friction time scale like the one employed here or another scheme to treat the merging of galaxies that no longer inhabit a (detectable) subhalo \citep[e.g.][]{Campbell2015}, there are always parameters involved that have historically been hard to constrain. Using the projected correlation functions together with an estimator as described in the current work offers a solution to this long-standing problem. Furthermore, the high sensitivity of the correlation functions to the details of the satellite distribution confirms the need for a highly accurate satellite profile if one aims to predict galaxy clustering.

\begin{figure*}
\begin{center}
\includegraphics[width=1.0\textwidth]{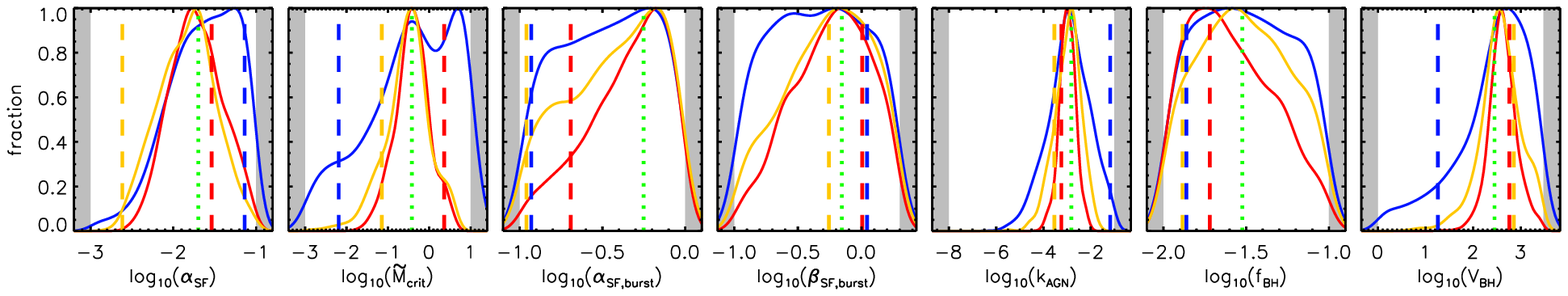} \\
\includegraphics[width=1.0\textwidth]{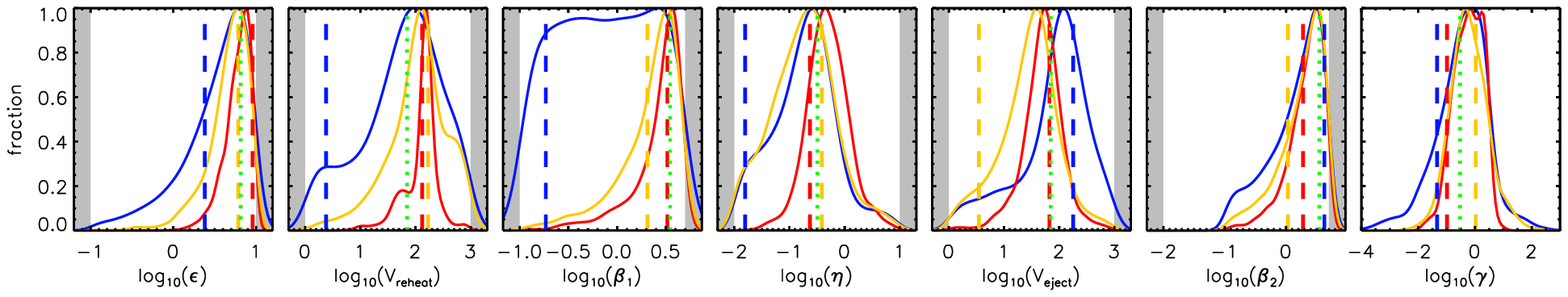} \\
\includegraphics[width=1.0\textwidth]{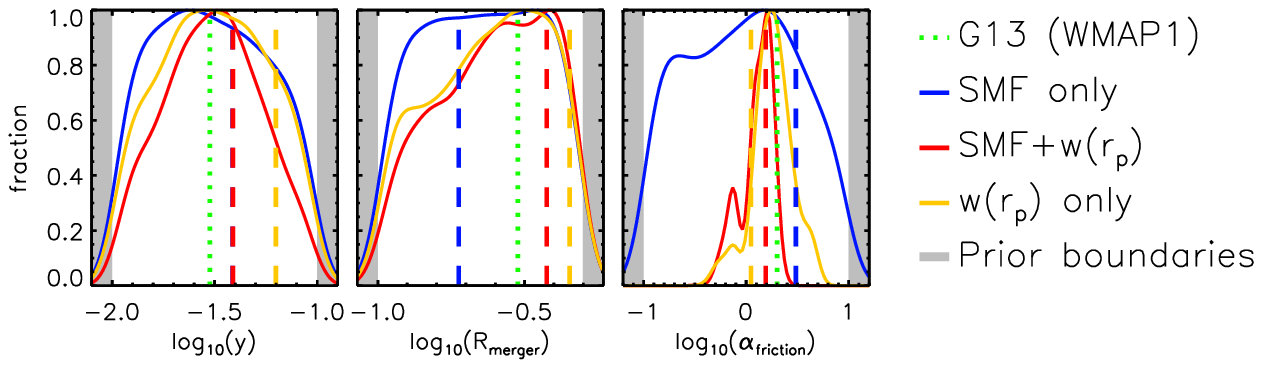}
\caption{The preferred parameter values in both models. The best-fit values are shown as dashed vertical lines. The dotted vertical line shows the parameters of the original G13 model, and the grey regions indicate values deemed non-physical, and which are therefore made inaccessible to the model. While shifts in $\alpha_\mathrm{friction}$ and $V_\mathrm{reheat}$ are the most important in order to obtain clustering predictions in agreement with the data, shifts in other parameters are typically needed to compensate for undesired changes in the stellar mass function. Wide likelihood regions typically indicate degeneracies, which are significantly reduced when clustering is used as a constraint, with respect to (only) using the SMF.}
\label{fig:params}
\end{center}
\end{figure*}

The parameter $V_\mathrm{reheat}$, on the other hand, shifts up by almost a factor of two with respect to G13 (from $70$ to $132\vunit$), increasing the effectiveness of supernova feedback on ISM gas as a function of halo mass. This similarly changes the satellite profiles, but while the decrease in $\alpha_\mathrm{friction}$ causes many orphans to merge away faster, flattening the distribution at all masses, the change in $V_\mathrm{reheat}$ shifts the relative distribution of high- and low-mass satellites. The indirect result of a more effective ISM heating is that the galaxy distribution profiles for low-mass galaxies flattened, while those for satellites with $M_* \ga 10^{10.77}\munitnoh$ are slightly steepened. The latter, undesirable effect is mitigated through changes in other parameters, mainly by lowering the effectiveness of AGN feedback.

\begin{figure*}
\begin{center}
\includegraphics[width=1.0\textwidth, trim=5mm 8mm 0mm 0mm]{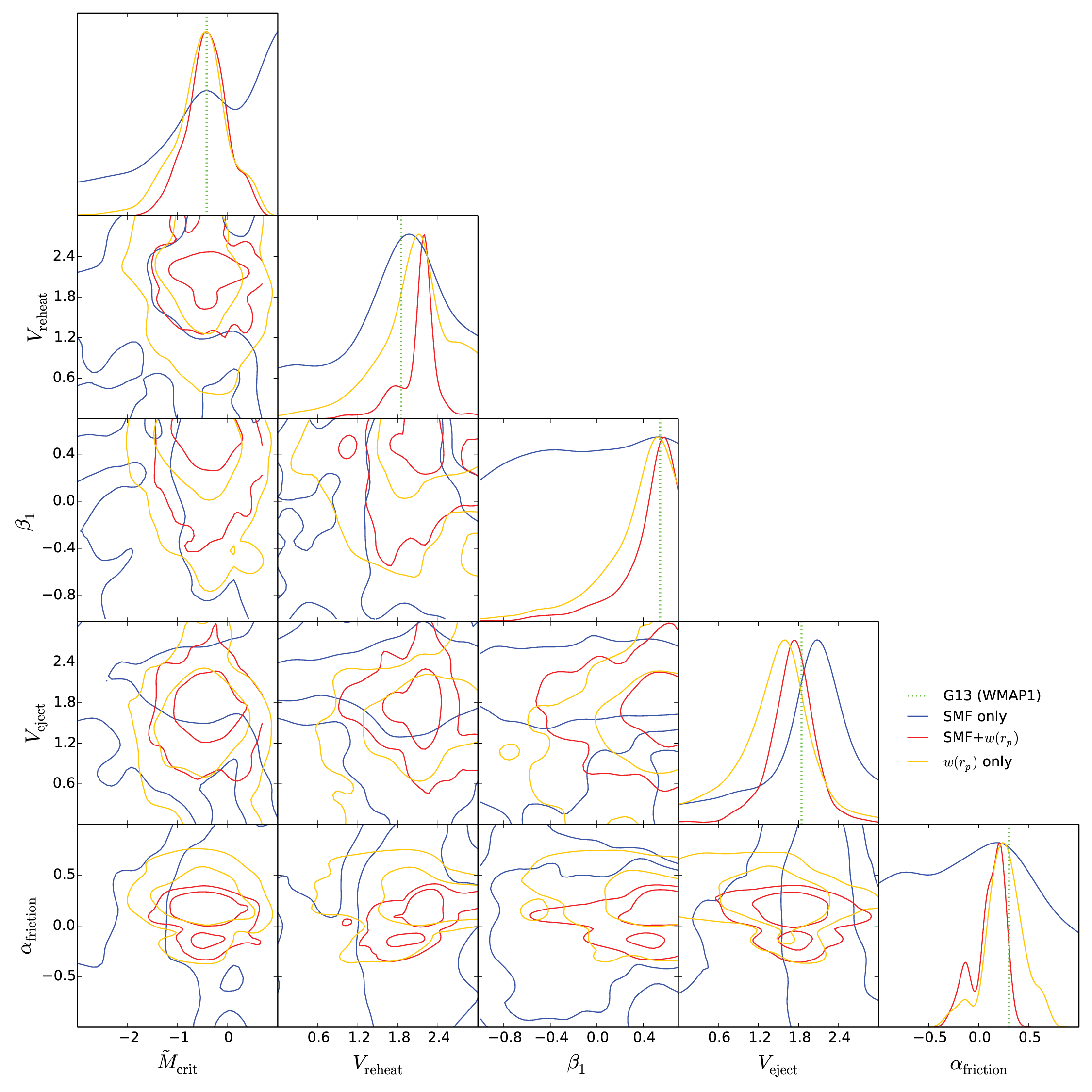}
\caption{Joint likelihood regions for the five parameters most influenced by the clustering constraints. They are $\tilde{M}_{\rm{crit}}$, which controls the threshold for star formation, $V_\mathrm{reheat}$ and $\beta_1$, which control the effectiveness of SN feedback heating ISM gas as a function of halo mass, $V_\mathrm{eject}$, which controls the effectiveness of stellar material being ejected from the galaxy by SN feedback as a function of halo mass, and finally $\alpha_\mathrm{friction}$, which controls the dynamical friction time scale of orphan satellite galaxies. Generally speaking, the likelihood regions of models that have the observed galaxy clustering imposed on them are narrower and more peaked than those of models that do not. Most notable are the changes in $V_\mathrm{reheat}$ and $\alpha_\mathrm{friction}$ when clustering data is included, which strongly favour a relatively small range in parameter space that is significantly offset from the input G13 value: these are the parameter shifts that are instrumental in bringing the models in agreement with the observed projected correlation functions.}
\label{fig:paramsubset}
\end{center}
\end{figure*}

The changes in $\alpha_\mathrm{friction}$ and $V_\mathrm{reheat}$ also impact the stellar mass function. First off, the decrease in the dynamical friction time scale causes the number density of galaxies above the knee ($M_* > 10^{10.5}\munitnoh$) to decrease. This counter-intuitive change in the SMF comes about because the cold gas in the merging satellites directly feeds the supermassive black holes in the centres of the central galaxies, increasing feedback from AGN and thereby the suppression of star formation. The increase in $V_\mathrm{reheat}$ has the same effect for $M_* > 10^{10.5}\munitnoh$, but increases the number density of galaxies below this mass scale. While this change in the SMF is quite intuitive, one might expect the overall clustering to increase/decrease with an increase/decrease in the number density of galaxies of the same mass, as the clustering scales with the HOD. While this does play a small role, it is the (normalised) galaxy distribution within each halo that is the real driving force behind the clustering predictions. Additionally, the effect of the HOD is partly mitigated by its normalization with $\bar{n}_\mathrm{gal}$, the mean number density of galaxies at some stellar mass. This also explains how a $w(r_\mathrm{p})$-only model that underestimates the SMF at any stellar mass could still lead to just as accurate clustering predictions as, say, SMF+$w(r_\mathrm{p})$.

Incidentally, these sometimes counter-intuitive changes in observable quantities caused by shifts in a single parameter, let alone the complicated interactions that can occur between different parameters changing at once, serve to demonstrate the importance of using a scheme like MCMC rather than attempting to set the values of a model -- be it SAM or hydrodynamical -- by hand.

The parameter changes in $\alpha_\mathrm{friction}$ and $V_\mathrm{reheat}$ with respect to the G13 parameters already produce clustering predictions that are very close to those of the SMF+$w(r_\mathrm{p})$ model. However, as they adversely affect the SMF, other parameter shifts are necessary to bring the SMF back into agreement with observation. We show the shift in all parameter values in Figure~\ref{fig:params}. We again indicate the results for all four models: the original G13 model (green dotted lines), the SMF-only model (blue lines), the $w(r_\mathrm{p})$-only model (orange lines), and the SMF+$w(r_\mathrm{p})$ model (red lines). Histograms indicate the (smoothed) Bayesian likelihood regions as derived from the full MCMC chains\footnote{For each of the three models we ran $120$ parallel chains with $3000$ steps each.}, while the vertical dashed lines indicate the best-fit values.

\begin{table*}
\begin{center}
\begin{tabular}{lcccccccc}
\hline
Parameter & & G13 (WMAP1) & & SMF only & & SMF+w($r_\mathrm{p}$) & & w($r_\mathrm{p}$) only\\
\hline
$\tilde{M}_\mathrm{crit}\,[\mathrm{M}_{\sun}\,\mathrm{km}\,\mathrm{s}^{-1}\,\mathrm{Mpc}^{-1}]$ & & 0.38 & & $6.5\times 10^{-3}$ & & 2.3 & & $7.1\times 10^{-2}$\\
$V_\mathrm{reheat}\,[\mathrm{km}\,\mathrm{s}^{-1}]$ & & 70 & & 2.39 & & 132 & & 169\\
$\beta_1$ & & 3.5 & & 0.19 & & 3.3 & & 2.0\\
$V_\mathrm{eject}\,[\mathrm{km}\,\mathrm{s}^{-1}]$ & & 70 & & 179 & & 66.4 & & 3.50\\
$\alpha_\mathrm{friction}$ & & 2.0 & & 3.06 & & 1.55 & & 1.11\\
\hline
\end{tabular}
\end{center}
\caption{The best-fit parameter values for each of the differently-constrained models for the parameters shown in Figure~\ref{fig:paramsubset}.}
\label{tab:paramvals}
\end{table*}

Without delving too much into details, this figure allows us to make several interesting observations. The best-fit values for the different models can be quite different from those of G13, even though the stellar mass and correlation functions it produces are not wildly different from those of our new models. As previously mentioned, the decrease in number density and the increase in clustering for high-mass galaxies in SMF+$w(r_\mathrm{p})$ caused by the shifts in $\alpha_\mathrm{friction}$ and $V_\mathrm{reheat}$ are mitigated by lowering the effectiveness of AGN feedback through a decrease in $k_\mathrm{AGN}$ and $f_\mathrm{BH}$. However, while in this case the best-fit parameters reflect these typical shifts, this is not the case for all high-likelihood parameter sets, i.e.\ the best-fit parameters do not necessarily fall near the peak of the likelihood region. For example, while the models that are constrained by the clustering data typically show the parameter shifts we just described, some high-likelihood parameter sets for the $w(r_\mathrm{p})$-only model (not shown here) leave $V_\mathrm{reheat}$ at its initial value and instead achieve a match to the clustering data by lower $\alpha_\mathrm{friction}$ only, and lowering it further than SMF+$w(r_\mathrm{p})$. Since it is more interesting -- and indeed more Bayesian -- to consider the typical solutions preferred by the model, rather than some best-fit set of parameters which may only perform marginally better than many others and do not inform us about the importance of its individual elements, we will only consider how the likelihood regions compare in what follows, both between the different models and to the G13 input values.

The likelihood regions for SMF-only are generally much wider than for the clustering-constrained models, which points to large degeneracies. This does not mean that the stellar mass function is unaffected by a shift in any of these parameters, but rather that a combination of shifts in other parameters can usually compensate for any undesirable changes caused. These degeneracies are already much less apparent when the clustering constraints are used, even if they are the only constraints. Indeed, by comparing the likelihood regions for $w(r_\mathrm{p})$-only and SMF+$w(r_\mathrm{p})$ (orange and red respectively), one can see that adding the stellar mass function as a constraint usually does not limit the parameter space traversed by the MCMC by much. As we used clustering data that was split into different stellar mass bins, this is not too surprising. It should be stressed, however, that modern semi-analytic models typically also take stellar mass or luminosity functions at higher redshift \citep[e.g.][]{Henriques2013,Henriques2015} as well other one-point functions such as colour information into account, which we have not done here. Regardless, it is clear that clustering can be a powerful and in some ways orthogonal addition to this list of constraints.

We further study the likelihood regions for a subset of the most interesting parameters in Figure~\ref{fig:paramsubset}. These are the five parameters for which the width of the likelihood region and/or the location of its peak relative to the G13 value changes most significantly when adding clustering constraints, meaning that the clustering constraints affect the limits on these parameters the most. It is clear from the relative widths of the likelihood regions that these parameters are more strongly constrained when the observed galaxy clustering is imposed. Besides showing wide likelihood regions, the SMF-only data often displayed multiple peaks, behaviour which is also reduced when clustering data is used. One notable exception is $\alpha_\mathrm{friction}$, although there the peaks do lie much closer together than for SMF-only.

The most essential parameters for lowering the small-scale clustering, $\alpha_\mathrm{friction}$ and $V_\mathrm{reheat}$, are also the ones that show the most interesting behaviour here. Not only are their distributions relatively narrow for SMF+$w(r_\mathrm{p})$ and $w(r_\mathrm{p})$-only, but their peaks are also clearly displaced from the input G13 values, favouring smaller dynamical friction time scales and an ISM reheating that is effective up to higher halo masses. While SMF-only also shows displays displaced peaks occasionally, there the G13 value is typically still in a region of high likelihood.

Finally, one interesting parameter to point out is $V_\mathrm{eject}$: this parameter is not noticeably more strongly constrained when clustering data is imposed, but does show a peak displacement from both SMF-only and G13, favouring a slightly lower effectiveness of supernovae ejecting material as a function of mass. This shift is mainly needed to counter the effect of the change in $V_\mathrm{reheat}$ on the abundance of low-mass galaxies, although other parameter combinations are able to serve a similar purpose (as evidenced by $V_\mathrm{eject}$'s still relatively wide likelihood region).

The best-fit parameter values for the five parameters shown in Figure~\ref{fig:paramsubset} are given in Table~\ref{tab:paramvals}.

\section{Summary}
\label{sec:summary}
We have developed a fast and accurate clustering estimator, capable of predicting the projected galaxy correlation function for a full simulated galaxy catalogue to within $\sim 10\%$ accuracy using only a very small subsample of haloes ($<0.1\%$ of the total sample). In this work, we have described our estimator and demonstrated its effectiveness for use in constraining parameter space for semi-analytic models of galaxy formation, using the \citet{Guo2013} version of the Munich SAM as a test case. Central to the success of our estimator is a new, highly accurate satellite profile, presented in equation \eqref{eq:preprofile}.

Our estimator determines the halo occupation distribution of galaxies in the subsample and fits a profile to the galaxy distribution within haloes as a function of halo mass, using these quantities in a halo model based approach to determine the galaxy clustering of the full sample. By being able to quickly predict the two-point galaxy correlation function for the first time while exploring parameter space, one can use clustering observations to limit the range allowed to the galaxy formation parameters of any SAM, adding constraints complementary to those of one-point functions typically used today, such as the stellar mass or luminosity function. As we have demonstrated, this substantially tightens constraints on parameters, and in some cases drives significant shifts in their preferred values. For suitable parameters, existing galaxy formation models nevertheless appear capable of reproducing well both clustering and abundance data for low-redshift galaxies. These results also imply that -- at least on the scales considered here -- the projected correlation function by itself may not be enough to constrain cosmology, as changes in galaxy formation physics can apparently compensate for using a set of cosmological parameters that significantly differ from current constraints.

For the G13 model tested here, the improved match to the correlation function is achieved mainly by significantly decreasing the time it takes for stripped (orphan) satellites galaxies to merge with their centrals (through a shift in $\alpha_\mathrm{friction}$), as well as increasing the effectiveness of SN feedback in heating the cold ISM gas as a function of halo mass (through $V_\mathrm{reheat}$). Both changes cause the galaxy distribution profiles within haloes to flatten, lowering the clustering on small scales. Other parameter shifts mainly serve to keep the changes in the SMF caused by the reduced time scales in check. The fact that the change in $\alpha_\mathrm{friction}$ has a stronger impact on the projected correlation function than on the measured satellite profiles implies that using clustering data as a constraint is likely the best way to find the right value for this type of parameter. This is an important result not just for semi-analytic models that use a dynamical friction time scale as employed in G13, but for any model that parametrises the merging of its orphan satellites.

While it is already accurate enough for our current application, several improvements could be made to the clustering estimator and/or its application. For example, the halo selection function can be calibrated to a higher accuracy than the $\sim 10\%$ accuracy we aimed for in the test case presented here, at the cost of a larger halo sample. This will improve the estimator's performance for high-mass galaxies especially, which will likely be important when cosmological parameters are allowed to vary as well as galaxy formation parameters. The higher computational resources could, for example, be offset by varying fewer model parameters simultaneously. Furthermore, the inputs to the model that are not derived from the sample galaxy catalogue, such as the input matter/halo power spectrum, could be improved by e.g.\ using higher-order bias terms (such as those recently presented by \citealt{Lazeyras2016}), or even a mass-dependent halo power spectrum measured from the base N-body simulation, which could be scaled with cosmology following \citet{AnguloWhite2010b}.

Relatively simple extensions of the methods set out in this paper will allow galaxy formation physics and cosmology to be constrained by abundances and clustering of galaxies at a variety of redshifts and separately for star-forming and passive systems. Galaxy-galaxy lensing could also be included among the constraints through a straightforward extension of our scheme. High-quality observational data are now (or soon will be) available in many of these areas, and it seems likely that requiring simultaneous and acceptable agreement with a single galaxy formation simulation will provide strong constraints not only on astrophysical but also on cosmological parameters. We will explore some of these topics in future work.

\section*{Acknowledgements}
The authors thank Joop Schaye for useful discussions and comments on an earlier version of the manuscript. MPvD also thanks Martin White for fruitful discussions on the Poisson distribution. The Millennium Simulation databases used in this paper and the web application providing online access to them were constructed as part of the activities of the German Astrophysical Virtual Observatory. This work was supported in part by the Marie Curie Initial Training Network CosmoComp (PITN-GA-2009-238356), by Advanced Grant 246797 "GALFORMOD" from the European Research Council, and by the Theoretical Astrophysics Center at UCB. This work used the DiRAC Data Centric system at Durham University, operated by the Institute for Computational Cosmology on behalf of the STFC DiRAC HPC Facility (www.dirac.ac.uk). This equipment was funded by BIS National E-infrastructure capital grant ST/K00042X/1, STFC capital grants ST/H008519/1 and ST/K00087X/1, STFC DiRAC Operations grant ST/K003267/1 and Durham University. DiRAC is part of the National E-Infrastructure.
\bibliographystyle{mn2e}
\setlength{\bibhang}{2.0em}
\setlength{\labelwidth}{0.0em}
\bibliography{PhDbib}

\appendix
\section{Fitting the satellite profile}
\label{fittingapp}
The projected correlation function is very sensitive to small changes in the satellite profile. It is therefore important not only that the functional form of the profile provide a good match to the simulation, but also that the fitting procedure be as unbiased as possible. In our case, fitting bias is an issue because when binning the satellite distribution for a small halo sample in a limited range of halo masses, many radial bins may be empty. Discarding these bins (or equivalently, assigning them infinite error) will bias the profile high, while including them with some finite error may bias the profile low. Additionally, some information is lost when considering the number of satellites in different radial bins as independent measurements. We find that these seemingly small effects can bias the estimated projected correlation function by about $10\%$ for all scales below $1\,\mathrm{Mpc}$.

To avoid these biases, we follow the following procedure. We interpret the measured satellite profile as a realisation of a series of Poisson distributions, with radially-dependent means $\mu$ that scale with the profile given in equation \eqref{eq:profile}. Specifically:
\begin{eqnarray}
\nonumber
\mu(r,\mathbf{p}) \!\!\!\!\!&=&\!\!\!\!\! \frac{N_\mathrm{sat}}{N_\mathrm{h}}\frac{n_\mathrm{sat}(r,\mathbf{p})}{\left<N_\mathrm{sat}\right>}\,\mathrm{d}^3r\\
\!\!\!\!\!&=&\!\!\!\!\! n_\mathrm{sat}(r,\mathbf{p})\,\mathrm{d}^3r\\
\label{eq:means}
\nonumber
\!\!\!\!\!&=&\!\!\!\!\! \frac{c}{4\pi b^3 r_\mathrm{vir}^3\Gamma\left[\frac{a}{c}\right]} \left(\frac{x}{b}\right)^{a-3}\exp\left\{-\left(\frac{x}{b}\right)^c\right\}\left<N_\mathrm{sat}\right> \,\mathrm{d}^3r.
\end{eqnarray}
Here we have explicitly shown the dependence of the profile on its parameters with $\mathbf{p}=\{a,b,c\}$. Note that we have slightly simplified the profile given in equation \eqref{eq:profile}: for a sufficiently large maximum scaled radius $x_\mathrm{max}$, any reasonable set of parameters gives $\gamma\left[\frac{a}{c},\left(\frac{x_\mathrm{max}}{b}\right)^c\right] \approx \Gamma\left[\frac{a}{c}\right]$. For our chosen value of $x_\mathrm{max}=5$, we find that this approximation is well justified. We do however still only consider satellites for which $x=r/r_\mathrm{vir}<x_\mathrm{max}$.

Next, we consider infinitesimally small radial bins, such that the number of satellites in each bin, $N_i$, is either zero or one. The likelihood function is then given by the product of the Poisson distributions at each radius, which we convert to a log-likelihood (exploiting the binary nature of $N_i$):
\begin{eqnarray}
\nonumber
\mathcal{L}(\mathbf{p}) \!\!\!\!\!&=&\!\!\!\!\! \prod_i \frac{\mu_i(\mathbf{p})^{N_i}\, e^{-\mu_i(\mathbf{p})}}{N_i!}\\
\Rightarrow \ln{\mathcal{L}(\mathbf{p})} \!\!\!\!\!&=&\!\!\!\!\! \sum_i \ln{\mu_i(\mathbf{p})} - \sum_i \mu_i(\mathbf{p})\\
\nonumber
\!\!\!\!\!&=&\!\!\!\!\! \sum_i \ln{n_\mathrm{sat}(r_i,\mathbf{p})} + \sum_i \ln{\mathrm{d}^3r_i} - \!\int\!\! n_\mathrm{sat}(r,\mathbf{p})\,\mathrm{d}^3r.
\label{eq:likelihood}
\end{eqnarray}
Since the profile is by definition normalised, only the first term has any residual dependence on the parameters $\mathbf{p}$. Ignoring constants, we therefore seek to maximise:
\begin{equation}
\sum_i \left[\ln{c}-3\ln{b}-\ln{\Gamma\left(\frac{a}{c}\right)}+(a-3)\ln{\frac{x_i}{b}}-\left(\frac{x_i}{b}\right)^c\right].
\label{eq:maxlikelihood}
\end{equation}
The scaled radii of all satellites in the halo mass bin are thus directly fed into the likelihood function, without radial binning. To maximise this term we utilise the derivatives of the log-likelihood function with respect to the different parameters, which are presented here for completeness:
\begin{eqnarray}
\nonumber
\frac{\partial\ln{\mathcal{L}(\mathbf{p})}}{\partial \ln{a}} \!\!\!\!&=&\!\!\!\! \sum_i \left[a\ln{\left(\frac{x_i}{b}\right)}-\frac{a}{c}\psi{\left(\frac{a}{c}\right)}\right],\\
\label{eq:likederivs}
\frac{\partial\ln{\mathcal{L}(\mathbf{p})}}{\partial \ln{b}} \!\!\!\!&=&\!\!\!\! \sum_i \left[c\left(\frac{x_i}{b}\right)^c-a\right],\\
\nonumber
\frac{\partial\ln{\mathcal{L}(\mathbf{p})}}{\partial \ln{c}} \!\!\!\!&=&\!\!\!\! \sum_i \left[1+\frac{a}{c}\psi{\left(\frac{a}{c}\right)}-c\ln{\left(\frac{x_i}{b}\right)}\left(\frac{x_i}{b}\right)^c\right].
\end{eqnarray}
Here $\psi(x)$ is the digamma function, defined as the logarithmic derivative of $\Gamma(x)$.

Any set of parameters must satisfy $0\!<\!a\!<\!3$, $b\!>\!0$ and $\!c>\!0$. We have tested that this method does indeed yield unbiased profile fits.

\bsp
\label{lastpage}
\end{document}